%% file: ms.tex
\documentclass[preprint]{aastex}
\newcommand{\simle}{\mbox{$\stackrel{<}{_{\sim}}$}}

\newcommand\arcdegg{\hbox{$^\circ$}}

\newcommand\etal{ {\em et~al.\/}\thinspace}
\newcommand\arcsecc{\hbox{$^{\prime\prime}$}}
\newcommand\asec{\hbox{$^{\prime\prime}$} }

\shorttitle{Mid-IR Sizes of YSOs}
\shortauthors{Monnier \etal}

\begin{document}

\title{Mid-infrared size survey of Young Stellar Objects: \\
Description of Keck segment-tilting experiment and 
basic results}


\author{J. D. Monnier\altaffilmark{1}, 
P. G. Tuthill\altaffilmark{2}, M. Ireland\altaffilmark{2}, R. Cohen\altaffilmark{3},
A. Tannirkulam\altaffilmark{1},  and M. D. Perrin\altaffilmark{4}
}

%\affil{Harvard-Smithsonian Center for Astrophysics,
%60 Garden Street, Cambridge, MA, 02138}

\altaffiltext{1}{University of Michigan, Department of Astronomy}
\altaffiltext{2}{University of Sydney}
\altaffiltext{3}{W. M. Keck Observatory}
\altaffiltext{4}{University of California at Los Angeles}
%\altaffiltext{5}{University of California at Berkeley}
%\email{jmonnier@cfa.harvard.edu, rmillan@cfa.harvard.edu}


\begin{abstract}
  
  The mid-infrared properties of
  pre-planetary disks  are sensitive to the temperature and flaring 
  profiles of disks for 
  the regions where planet formation is expected to occur.
  In order to constrain theories of planet formation, we have carried out 
  a mid-infrared ($\lambda=10.7\mu$m) size survey of young stellar
  objects using the segmented Keck telescope in a novel configuration.
  We introduced a customized pattern of tilts to individual mirror segments
  to allow efficient sparse-aperture
  interferometry, allowing full aperture synthesis imaging with higher
  calibration precision than traditional imaging.  In contrast to previous surveys on smaller
  telescopes and with poorer calibration precision, we find most
  objects in our sample are partially resolved. 
  Here we present the main observational results of our survey of
  5~embedded massive protostars, 25~Herbig Ae/Be stars, 3 T Tauri
  stars, 1 FU Ori system, and 5 emission-line objects of uncertain
  classification.    
  The observed
  mid-infrared sizes do not obey the size -- luminosity relation found at
  near-infrared wavelengths and a companion paper will provide further
  modelling analysis of this sample.  In addition, we report imaging
  results for a few of the most resolved objects, including complex
  emission around embedded massive protostars, the photoevaporating
  circumbinary disk around MWC~361A, and the subarcsecond binaries T~Tau, FU Ori
  and MWC 1080.
\end{abstract}



%\tableofcontents

\section{Introduction}

Theories of planet formation rely on observational estimates of
preplanetary disk initial conditions in order to make predictions.  By
connecting detailed and accurate disk initial conditions with the
observed diversity of exoplanetary systems, these theories promise to
explain how, when and where exoplanets form under a wide range of
circumstances.  Unfortunately, direct measurements of the initial
conditions are very uncertain since these regions remain largely
unresolved by conventional observing techniques. Indeed, disk theory
is still undergoing active revision and new measurements more often
reveal puzzling surprises rather than build confidence in
pre-existing models.

Pioneering near-infrared observations of young stellar objects (YSOs)
using long-baseline interferometry \citep{rmg1999a,millangabet2001,tuthill2001}
found disk emission sizes that were much larger
than expected from the conventional disk theory.
This has been successfully interpreted in terms of an optically-thin
cavity surrounding the star, in contrast to previous assumptions of 
an optically-thick disk extending to stellar surface \citep{monnier2002a,natta2001,monnier2005}.
While most work has been done for the intermediate mass Herbig Ae/Be stars,
this general pattern is seen even for the young sun analogues, the T Tauri stars
\citep{akeson2005b,eisner2005,ppv2006}.

The introduction of this large inner cavity to explain the near-infrared
sizes radically changed the inner disk structure in models of these systems, requiring a 
``puffed-up inner rim" and even a shadowed region behind the rim in some models
\citep{dullemond2001}.  
In order to probe the disk structure and surface temperatures at these larger radii,
we need to observe at longer wavelengths.  Mid-infrared size measurements 
using interferometric techniques promise to probe the terrestrial-planet forming
zones around young stars, characterized by temperatures around 300K.  Until recently,
most attempts to resolve YSO targets in the mid-IR have failed due to lack of angular resolution 
\citep[e.g.,][]{liu1996}.

Hinz and collaborators have resolved a few targets using a novel nulling interfometer BLINC
on a single 6.5m-aperture telescope \citep{hinz2001, liu2005,liu2007}, although most
targets were unresolved.  Clearly, longer baselines will be needed to
increase the angular resolution; the ISI 
interferometer was the first long-baseline interferometer to
resolve thermal emission from a YSO \cite[Lkh$\alpha$101;][]{tuthill2002}.

A radical technological advance was recently made by the MIDI
instrument on the VLTI interferometer, measuring seven mid-infrared
sizes of Herbig Ae/Be stars \citep{leinert2004} and even taking a
spatially-resolved spectra across the silicate feature
\citep{vanboekel2004}.  With baselines ranging from 10m to 200m, the
VLTI-MIDI system \citep[and also the Keck Interferometer Nuller;][]{stark2008} is poised to revolutionize studies of YSO disks in
the mid-infrared. 

In this article, we describe an experiment employing one of the
world's largest single-aperture telescope, the Keck-1 telescope.  We
have optimized the calibration precision of our survey by programming
the individual mirror segments of the Keck to create multiple
non-redundant interferometric arrays.  In this method, we allow
precise calibration of changing atmospheric conditions while
instantaneously sampling full 2-dimensional sky angles, allowing
detection of disk elongations and asymmetries with unprecedented
precision.  Furthermore, we can also survey sources very quickly,
allowing a large sample to be observed in only a few nights.  Our
study complements longer-baseline observations since ``short''
baselines are needed to distinguish between large-scale disk emission
and emission from the compact ``puffed-up inner rim''
\citep{vanboekel2005,tannirkulam2007}.

Here we report the full results of our diffraction-limited size survey
of bright young stellar objects, finding most of them to be
resolved and a few to be elongated.  In addition to our imaging results, we can investigate
the mid-infrared size -- luminosity relations for YSOs in detail for the first time.  
Unfortunately, the recent
decommissioning of the only mid-infrared camera on the Keck telescope
marks the end of this short-lived but exciting experiment.

\section{Observations}
\subsection{Description of segment-tilting experiment}
\label{keckseg}
\label{segmenttilting}

While mid-infrared imaging with 10-m class telescopes is often
considered ``diffraction-limited," precise measurements of
partially-resolved objects are difficult due to changes in seeing
between observing a target object and a Point Spread Function (PSF)
calibrator.  In order to circumvent these difficulties, we 
designed a novel interferometer experiment whereby individual segments
of the 36-segment Keck telescope were reconfigured to form four interferometric
arrays.  Since each segment (1.8m across) is much smaller than the typical
coherence length of the atmosphere (r$_0=$5-10m at $\lambda=$10$\mu$m)
the results are more robust to variations in seeing conditions.  The general
approach of apodizing a telescope for interferometry is known as
``aperture masking'' and the technique has well-established advantages over speckle interferometry and adaptive optics for bright targets.  
We refer the interested reader to
the pioneering work of \citet{haniff87} and 
the more recent implementation on Keck for further discussion
\citep{tuthill2000}.

\begin{figure}[hbt]
\begin{center}
\includegraphics[angle=90,width=7in]{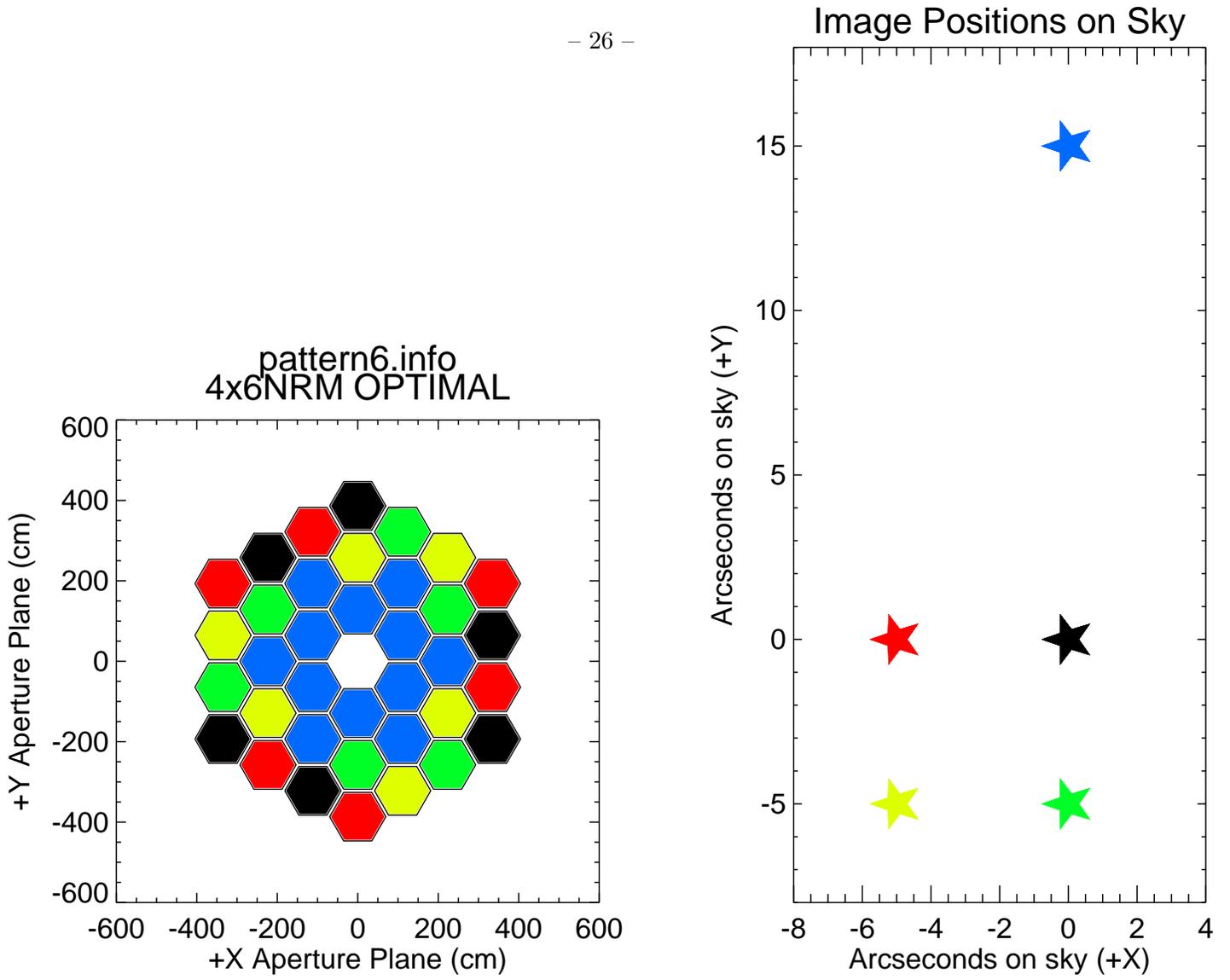}
\figcaption{\footnotesize 
{\em (left panel)} a. This figure shows the 36 segments of the Keck telescope, color-coded by the tilt imposed using the Keck Active Control System (ACS) for 
``pattern 6'' described in \S\ref{keckseg}.
 {\em (right panel)} b. This figure shows the image plane locations associated with the tilts
color-coded in panel (a). The 10\asec X 10\asec array detector can simultaneously record
fringe patterns of the bottom four image plane locations, while the top image is off the chip.
\label{pattern6}}
\end{center} 
\end{figure}

In this work, we formed five interferometric sub-arrays by tilting the
Keck primary mirror segments in a customized pattern.
Figure~\ref{pattern6}a shows the Keck segment map color-coded by
the applied tilt in a pattern known as ``pattern 6'' (while other tilt patterns were available, only pattern 6 was used in this work).  These tilts caused
the stellar image to split into 5 separate images as illustrated in
Figure~\ref{pattern6}b.  Four of the images (each with light from 6
segments) were arranged in a 5'' $\times$ 5'' square which fit
perfectly on the 10'' $\times$ 10'' field-of-view of the Long
Wavelength Spectrometer (LWS) camera on Keck-1, a 128$\times$128 array
with 0.83\asec platescale \citep{campbell2004, perrinthesis}.  
The four images were used for science and
were measured during data acquisition.    The segments contributing to each
of the four ``science" arrays were arranged in a non-redundant pattern
so that the fringe visibilities of all baseline pairs could be extracted
using Fourier analysis \citep[see discussion of redundancy in aperture
masking interferometry in][]{tuthill2000}. Because not all segments could be incorporated into four non-redundant patterns, a fifth image was created using light from the  12 unused segments and this light was projected
15\asec away from the detector.

This approach to aperture masking has a few significant advantages
over those employing opaque masks placed in the telescope pupil plane
(as done for all previous aperture masking work).  Firstly, since
light from 24 of the 36 segments are used for science, the collecting
area of the Keck primary is efficiently used (compare to near-infrared
Keck aperture masking which passes 1-10\% of the incident flux).
Secondly, each of the four sub-arrays generally has different (u,v)
Fourier coverage, which is essential for high fidelity aperture
synthesis imaging.  Indeed, we
have already presented high-resolution images of dusty evolved stars
\citep{weiner2006,ireland2007}, Wolf-Rayet stars \citep{rajagopal2007}, and one YSO \citep{monnier2008a} 
with unprecedented angular resolution.
Thirdly, tilting segments is operationally
straightforward and does not require any hardware modifications -- we
expect this technique to find application for the next generation of
segmented telescopes, e.g., Gran Telescopio Canarias,
Cornell Caltech Atacama Telescope, and the Thirty Meter Telescope.

\subsection{Segment-tilting procedure}
\label{procedure}
The segments were tilted by sending offsets to the 108 actuators which
support and control the Keck primary mirror.  For all segments in a
subarray, offsets were calculated so that incoming light was reflected
to the desired off-axis location in the telescope image plane. Note
that in addition to the tilts, a large piston offset was needed to be
applied to the segments in order to preserve phase coherence at each
of the four new pointing origins.  In essence, this extra piston term
was necessary so that each subset of segments conformed to a new
parabolic surface.  These piston offsets can amount to
$\sim$100$\mu$m, critical to correct so that interferometric fringes
fall within the coherence envelope of the light (typical narrow-band
filters have $\frac{\lambda}{\Delta\lambda}\sim7$).

In normal Keck operation, offsets between mirror segments are
continuously monitored and maintained by the Keck Active Control
System (ACS), employing capacitive edge sensors.  However, our
experiment required the edge sensors to operate with a lower gain
setting due to the much larger gaps between segment edges encountered
here. Unfortunately, the sensor response was not well-calibrated under
these conditions, and so the actual tilts and pistons realized in
practice only approximate the desired configuration (typical errors
were 5\asec).  Fortunately, the introduced errors were found to be
mostly corrected by employing so-called ``focus mode" to the primary.
``Focus mode" approximates telescope focus by changing the overall
curvature of the Keck primary mirror surface.  It is a well-known
characteristic of the Keck ACS system that changing sensor gain
settings introduce focus-mode error (Richard Cohen, private
communication).

\begin{figure}[hbt]
\begin{center}
\includegraphics[angle=90,width=6in]{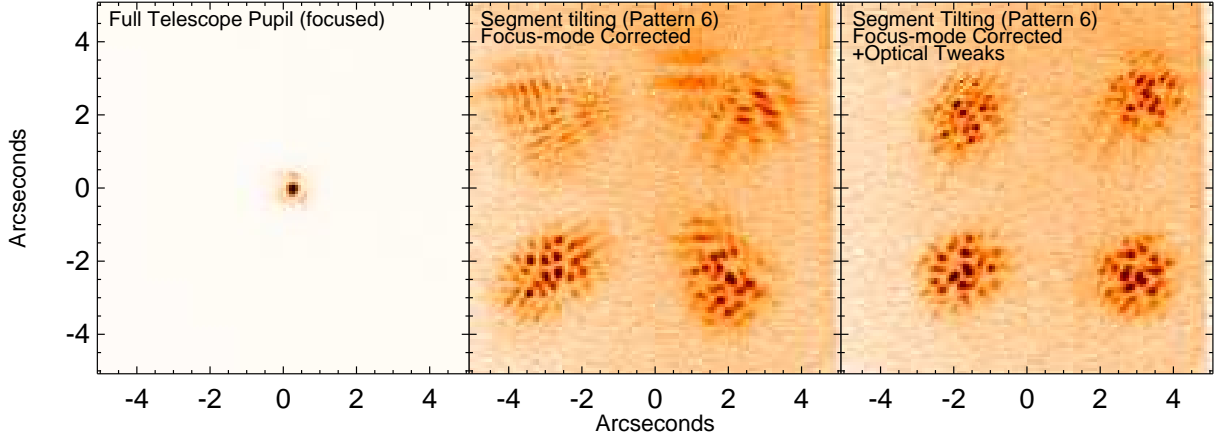}
\figcaption{\footnotesize 
{\em (left panel)} a. This figure shows a single frame (90~ms exposure) 
of a calibrator source using the full pupil and after careful focusing.  Although this image is
``diffraction-limited,'' there are significant variations of this PSF with time and seeing conditions and the first Airy ring is usually distorted by phase aberrations.
{\em (middle panel)} b. After applying the tilts and pistons for pattern 6 (see Figure~\ref{pattern6}), we see the light from the star is split into four patterns of overlapping fringes.  This image was taken after correcting for ``focus mode'' introduced when changing sensor gain setting (see discussion in \S\ref{procedure}).  Note the large residual tilt errors on some segments in the top of the image frame.
{\em (right panel)} c. The tilt errors from the middle panel were analyzed at the beginning of the observing night and perturbative ``optical tweaks'' were applied to the segment actuators.  The resulting image quality is much improved, showing four interference patterns with well-aligned segments.
The intensity patterns are all displayed with a linear scale.
\label{tweaks}}
\end{center} 
\end{figure}

After correcting focus-mode errors, there were still residual tilt
mismatches that affected some segments greatly.  Clearly,
if the light patterns from individual segment do not overlap well, the
fringe power will be reduced and become sensitive to small changes of
mirror figure (say, as a function of elevation).  In order to optimize beam
overlap, we applied a final correction using ``optical feedback.''  That
is, we took a set of data at the beginning of each night using our
best focus-mode alignment and analyzed the resulting fringe patterns to
estimate correction terms for each segment.  The algorithm we used
employed phase slopes in the Fourier transform of the speckle patterns
and would not have worked in the case of a redundant array of
segments.  We then applied these small corrections as a perturbation
to the existing ACS offsets.  This entire procedure took approximately
10 minutes and resulted in excellent
alignment, with individual segments overlapping with $\sim\frac{1}{10}$\asec 
precision, approximately 0.1 $\frac{\lambda}{D_{\rm segment}}$. See Figure~\ref{tweaks}
showing short-exposure fringe images before and after the final
``optical tweaking."

It was found that the segment alignment could change by up to
$\sim$0.5" as a function of telescope elevation and other operational
variables (e.g., temperature), necessitating re-calibration.  In
general, we solved this by having separate actuator calibration
``snapshots'' as a function of elevation and using calibrator stars
near to each target in time and elevation angle.  For bright objects
one can use post-processing to measure the tilts of each segment after
the fact (akin to the ``optical tweak'' algorithm discussed above) and
apply correction factors.  We did not use this extra calibration step
in this paper since the correction is signal-to-noise dependent and
many of the YSOs were too faint for robust implementation.

\subsection{Data collection methodology}

We employed standard chop-nod observing methods to minimize noise from
fluctuating background.  The secondary mirror was chopped at 5~Hz with
a 10\asec throw along the telescope azimuth direction.  In general,
each saved data frame had a total integration time of 90~ms and we
collected approximately 350 individual chop-nod sets.  We sometimes
employed longer integrations and slower (2.5 Hz) chop cycles for faint
targets, and also collected more chop-nod sets to improve
signal-to-noise ratio.  The 90~ms integration times were sufficient to
essentially ``freeze" any fringe blurring due to atmospheric
turbulence at these wavelengths, although telescope wind-shake can
still cause calibration problems; fortunately, wind-shake was minimal
during the observations presented here.  Each target was observed two
independent times, bracketed by observations of point-source
calibrators located nearby in the sky.  Table~1 contains the complete
observing log, including the target names, dates, times, and
calibrators used.  We observed all of our targets in only one filter,
the 10.7$\mu$m filter (passband 9.92-11.47$\mu$m).

\subsection{Data reduction}
The bulk of the data analysis processing was carried out by the same
aperture masking code developed for the near-infrared version of the
Keck masking experiment, which is best described in \citet{monnier99}
and \citet{tuthill2000}.  The only major modification to the data pipeline
was a change meant to address significant camera noise in the power spectra of 
our short exposure images.  Here we used the nod sequences (which did not have any star light) 
to estimate a contemporaneous bias power spectrum.
This bias subtraction proved necessary and key to extracting reliable results for the faintest
targets in our sample.

The overall observing procedure was introduced in the
first segment-tilting publications
\citep{weiner2006,rajagopal2007,monnier2008a}.  Essentially, each
subarray pattern (field-of-view 3\asec) was analyzed using Fourier
techniques, resulting in fringe visibilities for each baseline and
closure phases for all triangles.  These values were calibrated using
point-source reference stars.  Some objects are very resolved and the
shortest inter-segment spacing is approximately 2 meters long, meaning
we are missing a lot of short-baselines.  In order to make up for
this, we also analyzed the same data frames using baselines shorter
than 1.8~meters, which correspond to sampling the many redundant
baselines within a segment.  These short baselines are useful when
reconstructing a wide binary companion during apertures synthesis
imaging and helps to constrain information on the percentage of the
light coming from a large-scale ``halo'' for some targets.

The measurements from the four simultaneous (inter-segment) patterns
and the results of the ``short-baseline'' (intra-segment) measurements
are merged together and used in subsequent analysis.  We used
observations of the known binaries MWC~1080 and SVS~20 to validate our
position angles and plate scale.  The calibrated data are saved in the
OI-FITS data format for optical interferometry \citep{pauls2005} and
are available upon request.

\subsection{Analysis of Systematic Errors}
\label{systematics}
Since we expected some of our target sample to be``unresolved'' it was
critical to establish reliable upper limits in those cases.  This
required an extensive analysis of systematic errors, as was recently
done for near-infrared aperture masking data \citep{monnier2007a}.

The best method for reliable error estimation is to obtain multiple
{\em independent} datasets for each object.  This requirement was
a driving consideration in our observing strategy and we obtained
multiple observations for all our targets as can be seen in Table~1.
By analyzing the consistency between the multiple independent
measurements we can provide robust error analysis.

\begin{figure}[hbt]
\begin{center}
\includegraphics[angle=90,width=7in]{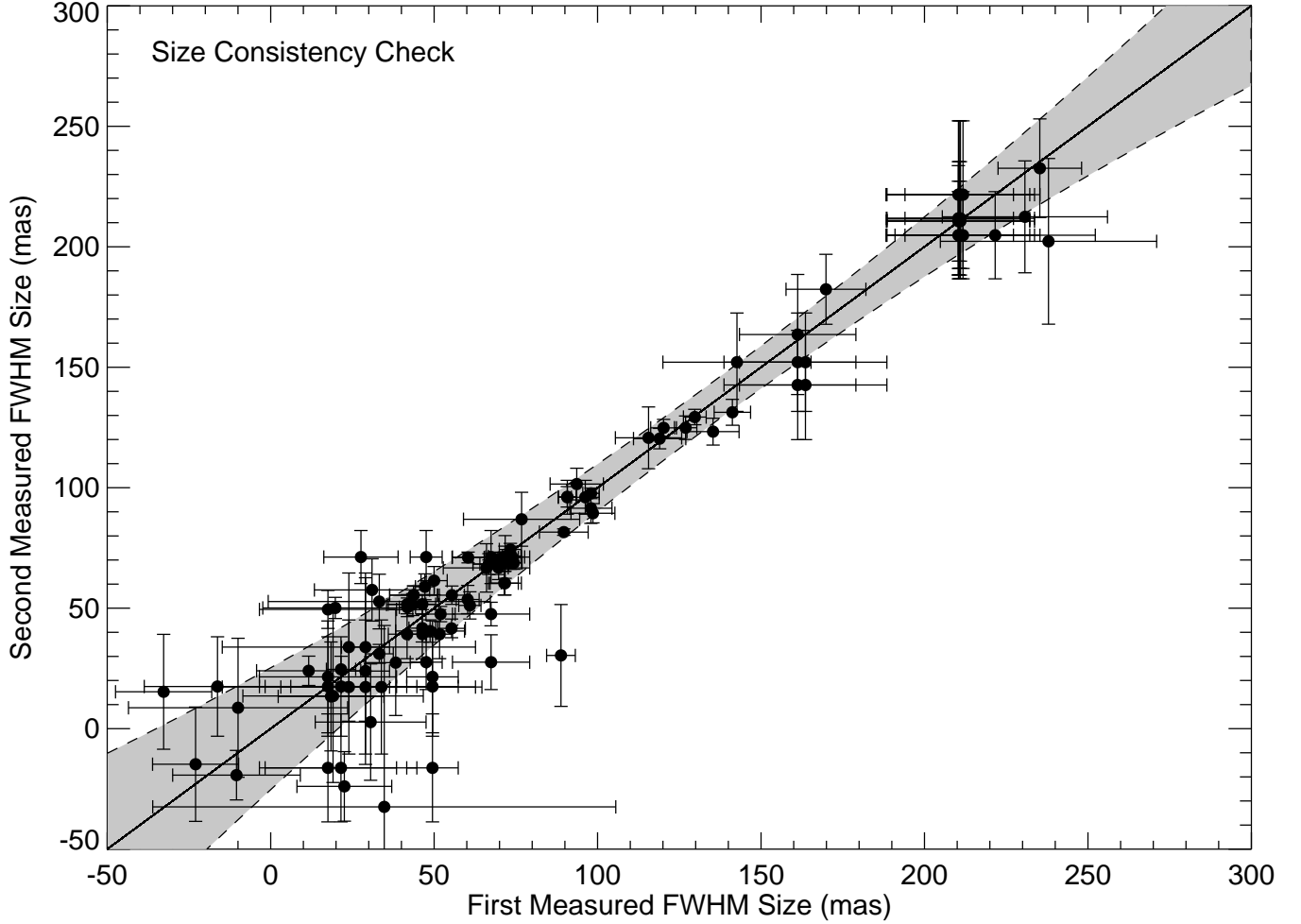}
\figcaption{\footnotesize 
Nearly all the targets of the Keck segment-tilting experiment were observed multiply times under independent observing conditions -- this was a core design principle of our observing methodology.
Here we plot the measured size (Gaussian FWHM) at one epoch versus another in order to estimate systematic errors
as a function of the target size.  Here we see extremely good self-consistency within 
reported errors and the typical 1-sigma confidence limits for the mean diameter 
are shown in the central grey-shaded region. We can reliably (with 2-sigma
confidence) estimate partially-resolved objects down to 
a full-width at half-maximum of $\sim$35~milliarcseconds, about 10$\times$ smaller than the 
formal diffraction limit.
\label{consistency}}
\end{center} 
\end{figure}

Figure~\ref{consistency} Shows a consistency plot for all the data of
the Keck segment tilting experiment using Pattern~6 -- this includes
many objects not part of this YSO survey but still useful for the
analysis of systematic errors.  Here we graph the size measured for an
object at one time against the size measured for the same target at
another time (and nearly always using an independent calibrator
sequence).  To create this graph we removed known binaries (MWC 1080,
T Tau, FU~Ori) and the error bars were estimated by analyzing the
4-subarrays of pattern 6 using bootstrap sampling techniques.  Based
on the variation in measured diameter as function of mean diameter, we
can quantitatively estimate our 2-sigma detection limit as FWHM
35~milliarcseconds -- our confidence limits as a function of mean
diameter are also included in Figure~\ref{consistency}.

As was found in \citet{monnier2007a}, the calibration precision is
non-linear in the sense that the fractional error is very large for
small sizes  while being greatly reduced for
larger sizes.  
Our final fitting results (both 1-D and 2-D) used bootstrap sampling of the
different epochs and sub-arrays to capture calibration and systematic errors.
In the case that two independent measurements disagreed
dramatically, we removed this target from our sample.  As can be seen
in Figure~\ref{consistency}, this rarely occurred and did not
significantly impact our sample size.

\subsection{Aperture synthesis imaging}
\label{imaging1}
Aperture synthesis imaging can be carried out for the most resolved targets, useful for detecting faint diffuse emission and for detecting
binary companions.  We used the publicly available BSMEM image reconstruction software
\citep{buscher1994,lawson2004,lawson2006} for aperture synthesis
imaging.  This program was discussed recently in \cite{monnier2008a} and was validated in more detail against the MACIM \citep{macim} algorithm in \citet{zhao2008}.  BSMEM uses the maximum entropy method \citep{mem83,mem86} and is similar to the VLBMEM program 
\citep{sivia87} extensively used in the near-IR Keck masking project  
\citep[e.g.,][]{monnier99a, tuthill99, tuthill2000}.
While most young stellar objects were only partially resolved here, a few 
sources were large enough (on the sky) to be imaged  and these are discussed in
\S\ref{imaging}.

\section{Results}

Table~2 contains the complete source list for this work, including
5~embedded massive protostars, 25~Herbig Ae/Be stars, 3 T Tauri stars,
1 FU Ori systems, and 5 emission-line objects of uncertain
classification. We have also included other important characteristics of our sample, including
V, J, H, and K magnitudes, our measured 10.7$\mu$m flux within aperture of 3\asec diameter, 
the IRAS 12 and 25$\mu$m flux, and
 notation regarding the presence of a close binary companion (based on the literature).   
Subsequent tables and figures maintain the
groupings and target orderings as presented in Table~2.

\subsection{Visibilities}
Figure~4 shows the squared-visibility data for all our targets as a
function of baseline in meters.
In this figure, 
we have performed (u,v) averaging on
scales of 0.45 meters.  This averaging is done in the (u,v) plane so
 baselines are only merged when sharing similar lengths and position
angles.  Here we did not average together results from the different
independent observations, but rather overplotted each separate epoch
detailed in Table~1.  Thus, the observed scatter represents the
combination of observing errors plus any position-angle dependency of
the source structure.  For instance consider the visibilities of
v892~Tau and MWC~349A which vary for
the same baseline length outside the error bars -- indeed, these two
objects are extremely asymmetric.  Also consider that MWC~1080 and T~Tau are both 
 well-known sub-arcsecond binary stars -- this is obvious in the
scattered visibility data.

Figure~5 shows the squared-visibility as a 2-dimensional function of
(u,v) coordinates, a useful presentation format for detecting elongations.  
For instance, the extreme asymmetry for
MWC~349A and v892~Tau, inferred from the 1-D visibility curves, is quite evident here.  
Also, the large ``halo'' structure around
MWC~361A, indicated by the sharp drop of visibilities as short
baselines and then a plateau, can be easily spotted.  The binary signatures in T~Tau, FU~Ori,
and MWC~1080 are evident too, although it is difficult to interpret since
the sinusoidal patterns are under-sampled, requiring analysis through
imaging or model-fitting to extract the binary component separation.

\subsection{Fitting Results}
\label{fitting}
We have characterized the emission sizes of all targets by fitting the
visibility data with both a circularly-symmetric (1-D) Gaussian model
and an elliptical (2-D) Gaussian model.  In both cases, we also allowed for a
large-scale ``halo'' component, similar to the model used for
near-infrared fitting of YSO visibilities in \citet{monnieriota2006} which was 
motivated by earlier work \citep[e.g.][]{leinert2001}.

Table~3 contains the complete fitting results from our work.  Error
bars were estimated through bootstrap sampling of the multi-epoch data
as well as from the different sub-arrays of ``Pattern 6.''  For each
fit (1-D and 2-D) we also report the best fit $\chi^2_\nu$ normalized
by degrees of freedom.  Thus, if the $\chi^2_\nu$ value is greater
than $\sim$1 then we expect this model to be an incomplete description
of the data. For instance for MWC~349A, the $\chi^2_\nu$ is 4.0 for the
circular-symmetric fit but decreases to 1.1 for a 2-dimensional
Gaussian -- we already discussed above that this object was clearly
asymmetric. In most cases when the $\chi^2_\nu$ is $\simle$1 for the
1-D fit, then the 2-D fit is only marginally better and that
there is no statistically-significant ellipticity detected.  Some
targets have high $\chi^2_\nu$ even for the 2-dimensional Gaussian
model, indicating a more complicated object.  This is true for all the
embedded YSOs which are generally the most resolved -- these objects
are ideal targets for aperture synthesis imaging in the next section
\S\ref{imaging}.

\subsection{Imaging}
\label{imaging}
While most of the analysis in this work and the companion paper relies
on the size-fitting results, we also report imaging results for the most
resolved targets.
We used the BSMEM image reconstruction program (see \S\ref{imaging1}) on
all targets and inspected the results.  In Figure~\ref{images}, we collected the images of all the
objects that showed structures beyond a simple, partially-resolved
Gaussian profiles and/or were larger than 70~mas in size.  
Note there are residual imaging artifacts at the level of a few percent of the
peak, especially evident for the binary sources MWC~1080, T~Tau and FU~Ori.
Notes on individual objects follow.

%Note\subsubsection{Comments on the imaging targets}

AFGL~490 has extension to south.

Mon~R2~IRS~3 shows nebulosity along a N-NE to S-SW axis and a companion at
 $\rho=836$~mas, $\theta=16.7\arcdegg$, similar position to the companion 
first reported by \citet{mccarthy1982} of separation 870~mas at PA 13.5$\arcdegg$.
% flux ratio is 32

AFGL~2136 has a  slight extension to west-southwest that appears in two independent sets.

AFGL~2591 emission has slight extension to west.

S140~IRS1 shows significant nebulosity to the South.

AB~Aur is a prototypical Herbig Ae star which has been claimed to be
elongated nearly 2-to-1 along PA~30$\arcdegg$ by \citet{liu2005}.  We
rule out this large level of elongation (supported by 4 separate
observations over 3 different observing nights), although we do find a
slight extension along this position angle in our image.  Overall the emission is
circularly symmetric, consistent with a nearly face-on geometry and in agreement with the
conclusions of \citet{marinas2006}.

v892~Tau was resolved into a circumbinary disk and was discussed in \citet{monnier2008a}. 
Here, we reproduce the image.

LkH$\alpha$~101 shows a slight elongation along North-South direction, similar to 
reported asymmetry in the near-infrared \citep{tuthill2002}. Note we do not
have fine enough angular resolution to resolve the dust-free cavity in the center reported
by these authors.

R~Mon is resolved approximately as much as AB~Aur, showing mostly symmetrical emission.

Z~CMa is a well-known close binary ($\sim$100~mas) recently imaged by
\citet{rmgzcma2002}. Here, the mid-infrared emission appears extended
towards position angle $\sim$128$\arcdegg$, which is the direction of
the FU-Ori type companion. This extension could represent emission
from a circumbinary disk or direct IR emission for the FU~Orionis
component to this system.  Most likely, the bulk of the mid-IR emission is
from the Herbig Be component of the system and not the FU Ori object.

MWC~349A is very elongated and the imaging shows a symmetric structure oriented along PA 95$\arcdegg$,
very similar to the PA 100$\arcdegg$ reported in the NIR \citep{danchi2001}.

MWC~361A was found to have a large ``halo'' containing 45\% of the
flux at 10.7$\mu$m.  This halo was over-resolved on our shortest
baselines.
Our shortest baseline
data suggested the halo was elongated north-south on arcsecond scales
and this is confirmed by independent (standard full-aperture) LWS
images taken at 11.6$\mu$m (SiC filter) and 17.65$\mu$m. These images
are shown in Figure~\ref{perrin_images}; a full description of these
observations and the data reduction appears in \citet{perrinthesis}.
Note the N-S elongation of the extended emission  matches the orientation of the
binary orbit of MWC~361A measured by
\citet{monnieriota2006}, strongly suggesting this halo is the remnant of a 
circumbinary disk.  It is remarkable that the mid-IR disk emission  
is $>$20$\times$ bigger than
the semi-major axis of the binary, indicating a huge disk gap in this system much larger than can be produced by disk clearing through dynamical interactions with the inner binary.
Given the early spectral type of the primary star in this system, 
we suggest that photoevaporation of the circumbinary disk is well underway.
%Strange!

MWC~1080 is binary with  $\rho=764$~mas, $\theta=-91.1\arcdegg$, comparable to the result of
\citet{leinert1997} of  $\rho=760$~mas, $\theta=-93\arcdegg$.
%ratio 13.4

The T~Tau companion to the south (which is itself a close binary) 
is detected at $\rho=639$~mas, $\theta=-170.9\arcdegg$
We attempted a more sophisticated analysis to look for evidence of the
tertiary to this system. Indeed, we find that that the visibility and closure phase data
can not be fit with a simple binary star \citep{skemer2008}.  
However, there are calibration issues regarding
field-of-view limitations caused by bandwidth-smearing that we have not corrected here.
This issue will be tackled in a future paper.
%flux ratio 13.

We detect a  companion to FU~Ori at separation 488~mas and PA of 
163.3$\arcdegg$ E of N, confirming the report by \citet{wang2004} of a potential companion
at separation 500~mas at PA 161$\arcdegg$.
As for T~Tau, we can not report a reliable flux ratio due to unresolved issues with 
%Intensity ratio is not reliable due 
bandwidth smearing effects that have not yet been corrected.
% flux ratio is 16.4

\subsection{Comparison to Literature}
Our results are generally in agreement with results from the BLINC
experiment by Hinz and collaborators \citep[][excepting the elongation
of AB~Aur]{hinz2001,liu2005,liu2007}, although we note most of these
workers' measurements were upper-limits.  We also compared our results
to the long-baseline interferometry VLTI data of \citet{leinert2004}
and others \citep{quanz2006}, finding that the Gaussian fits do not
always agree between the two methods (our sizes typically larger).
The size discrepancies can be understood if the long-baseline VLTI 
measurements are more sensitive to the properties of the hot inner wall while the short-baseline (single) Keck data can only probe the large-scale, outer 
flared disk \citep[see discussion in][]{vanboekel2005,tannirkulam2007,ppv2006}.
These observed differences appear to confirm the expectations from models that
mid-infrared emission comes from (at least) two distinct spatial scales
and we will need measurement from a wide range of scales in order to
reconstruct the actual disk temperature profile.

%Indeed, these differences
%illustrate why one must combine sizes measured from single aperture
%telescopes with data from long-baseline interferometers to accurately
%reconstruct the disk temperature profile.  

\section{Analysis}
\subsection{Size -- Luminosity Diagram}

A very tight correlation has been found between the near-infrared size
of a young stellar object disk and the luminosity of the host star for
Herbig Ae and late Be stars \citep{monnier2002a,monnier2005,ppv2006}.
This simple correlation arises because the NIR emission only comes
from the hottest dust near the evaporation front and the location of the
evaporation front is primarily sensitive to a single observable, the
central luminosity.

The mid-infrared emission from young stars is expected to come from up
to three regions: the hot inner wall located at the dust evaporation
radius, a thin surface layer on the disk
\citep[e.g.][]{vanboekel2005,tannirkulam2008b}, and perhaps a
circumstellar envelope (``halo'').  Thus, the mid-infrared size and
flux density will depend on multiple factors in addition to the
central luminosity, depending most sensitively on the dust size
distributions both vertically (setting temperature) and radially
(controlling disk flaring).  While it is beyond the scope of this
paper to investigate these effects, we will discuss them generally in the
context of the mid-IR size -- luminosity diagram.

Figure~\ref{sizelum} shows the mid-infrared size -- luminosity diagram
for the stars in our sample, constructed using the same procedure as
in \citet{monnier2002a}.  Our sample represents a factor of 3 times
increase in sample size over previous work and spans a large range of YSO
luminosity.  We used distances and luminosities from \citet{acke2004} when available. 
In this diagram, we find a crude correlation of size with luminosity 
showing disk
emission with characteristic temperatures of 250--900~K spanning about
5 orders of magnitude in luminosity.  While the previously-published
{\em near-infrared} size -- luminosity diagram shows only about a factor of 2
scatter at a given luminosity, this mid-infrared relation shows much
larger scatter, a factor of 5 or more.  The origin(s) of the increased
scatter and the breakdown in the size-luminosity relation will be the
subject of a companion paper (Tannirkulam et al., in preparation).
Here, we will only make general comments.

In \citet{tannirkulam2008b}, the large mid-infrared size difference
between two ``twin" Herbig Ae stars (AB~Aur and MWC 275) was explained
through the influence of small grains in the AB~Aur disk at radii of
$\sim$7~AU.  These grains absorb stellar radiation effectively,
heating up the disk and causing flaring.  The origin of these
small grains could be due to collisions between planetesimals or the
release of small grains as ice-cemented dust agglomerates break apart
at about $\sim$10~AU.  This example illustrates how variations between targets in
the {\em radial} dust size distributions can explain some of the
scatter in this diagram.  Other modifications to the dust distribution, such as dust growth and settling, can also cause significant changes in mid-infrared SED \citep{dullemond2004,dalessio2006}.

Another source of scatter is the presence of binaries.  We have indicated
close binaries in Table~\ref{table_sources} and in Figure~\ref{sizelum}.  
For instance, the emission from v892~Tau was recently imaged to
be from a circumbinary disk \citep{monnier2008a} and we see in
Figure~\ref{sizelum} that v892~Tau is indeed oversized for its luminosity.

We also note that the embedded YSOs all cluster near the top-right of the
diagram at high luminosities and low temperatures. This is expected since the
high optical depths toward these Class I objects mean their infalling
envelopes are still optically-thick and we do not expect to be able to
see into the warmer inner regions surrounding these protostars. Note
that these regions all show complicated, often bipolar, structures
(see Figure~\ref{images}) and that the sizes in this diagram refer only to the
central core of emission.

\subsection{Size vs IRAS color}

\citet{leinert2004} found some evidence that the physical emission
scale for the mid-infrared emission (in Herbig Ae stars) was
correlated with the IRAS 12-25 micron color, defined as
-2.5~log~F$_\nu$(12$\mu$m)~/~F$_\nu$(25$\mu$m).  This is sensible
since redder colors indicate cooler dust temperatures which naturally
arise farther from the central star.  Unfortunately, this diagram is
difficult to use when your target sample spans a large luminosity
range.  This is because the radius for dust at given temperature
depends on the root of central luminosity (see size-luminosity diagram
above).  Neither size-luminosity nor the size-color diagram can
simultaneously take into account both of these effects.
\citet{liu2007} also explored mid-IR size differences as a function of
the some SED parameters but his work also suffers from these same
limitations.

Despite these drawbacks, we have produced a similar diagram as a way
to present our basic results.  Figure~\ref{sizeiras}a shows our
measured Gaussian FWHM size (in milliarcseconds) vs. the IRAS 12-25
color.  We chose to use the distance-independent angular size instead
of the physical size here, just so that our quantities did not depend on the uncertain distance
estimates.
Within each grouping, we do see some correlations. The Herbig Ae stars do show
a correlation, with the striking result that nearly all the objects with colors
bluer than 0.25 were unresolved by our survey.  The embedded objects 
also might show a weak correlation.
Just as the diagram in \citet{leinert2004}, this size-color diagram does not
conserve a target's position if you take the same disk temperature profile 
and change the 
luminosity. 
We only present this data to illustrate some trends within the
sample, but caution against further use of this sort of 
diagram in subsequent work. 

In order to overcome the luminosity and distance-dependencies for this sort of diagram,
we introduce a new diagram in Figure~\ref{sizeiras}b.  Here plot the luminosity-normalized ring radius versus iras 12-25 color.  This is defined as the physical ring radius (AU) divided by the square-root of luminosity (in solar luminosities).  This quantity is both independent of distance and also naturally accounts
for the expected scaling of emission scale with luminosity (for similar disk temperature profiles).
Again, we see a significant correlation for the Herbig Ae stars, although other objects appear to
scatter in this diagram.  The location of each YSO system in this diagram is diagnostic of the
disk flaring and temperature profiles and this diagram will prove useful for future work.

In the next paper in this series, we will be carrying out a radiative
transfer study of YSO disks in order to understand the scatter in the
size-luminosity diagram and how the deviations correlate with 
observables such as stellar luminosity and the [12]-[25] micron color.
This new work will motivate improved diagrams to use for plotting basic
observational data and will explore the potential for using
luminosity-normalized sizes
(radius AU / sqrt Luminosity) and surface-brightness relations.

\section{Summary}

We have presented a large survey of mid-infrared sizes of young stellar objects
using the Keck Telescope. We can reliably resolve disks with size scales down to 
35~mas (5~AU at 140pc), roughly 8$\times$ smaller than the formal diffraction limit.  We have overcome the traditional calibration problems that have plagued previous surveys by reconfiguring the Keck segments into multiple non-redundant interferometric arrays.

We have presented a detailed description of the experimental methodology and 
validated our calibration precision.  We have measured the characteristic sizes of all our targets
using a Gaussian $+$ Halo model  and have presented aperture synthesis imaging when possible.
Our atlas of diffraction-limited imaging has revealed
the complicated emissions around a sample of Class-I embedded YSOs and also 
discovered the remarkable photoevaporating circumbinary disk around
MWC~361A.

Lastly, we have presented the most complete size -- luminosity diagram
for YSOs in the mid-infrared.  Notably, we found the mid-IR size --
luminosity relation shows a factor of 5-10 scatter for a given
luminosity, much larger than the tight correlation seen in the
near-infrared.  This large scatter can partially be understood in terms of the
influence of small dust grains that can be created in collisions between
planetesimals in these young planet-forming disks \citep{tannirkulam2008b}.  
Disk clearing by binary companions and 
emission from remnant dust envelopes may also contribute to this scatter.  
With the help of
a large grid of radiative transfer models, we will be fully exploring
how the mid-infrared sizes correlate with disk properties in the next
paper in the series.

While this particular experiment can no longer be carried out due to
the absence of mid-IR instrumentation on the single Keck telescopes,
we hope our method can be applied on current and future proposed
segmented telescopes, such as the Gran Telescopio Canarias, the
Cornell Caltech Atacama Telecope, and the Thirty-Meter Telescope.

\acknowledgments{
We have appreciated the involvement of C.H. Townes  in this project, especially for
his contributions to the initial work.
We give special thanks to Mark Kassis and Randy Campbell for superb observing
assistance and helpful discussions concerning experimental details.  We also 
thank Jayadev Rajagopal for his help at the telescope.
We acknowledge partial support from NASA KPDA grant
  1267021, NASA Origins grant NNG05GI80G, NSF-AST 0352728, and the
  NASA Michelson Fellowship program (MJI). This research has made use of the SIMBAD database,
operated at CDS, Strasbourg, France.
The data presented herein
  were obtained at the W.M. Keck Observatory, which is operated by
  Caltech, University of California and NASA.  WMKO
  was made possible by the financial support of the W.M. Keck
  Foundation.  The authors wish to recognize and acknowledge the very
  significant cultural role and reverence that the summit of Mauna Kea
  has always had within the indigenous Hawaiian community.  We are
  most fortunate to have the opportunity to conduct observations from
  this mountain.

{\it Facility:} \facility{Keck:I (LWS)} 
}

\bibliographystyle{apj}
\bibliography{apj-jour,LWSYSOs,HerbigSizes,WR140,Thesis,Review,iKeck,Review2,KeckIOTA,IONIC3,RS_Oph,v892tau}

%\begin{thebibliography}{apj-jour,CIT_6,Thesis}
%\end{thebibliography}

%% Generally speaking, only the figure captions, and not the figures
%% themselves, are included in electronic manuscript submissions.
%% Use \figcaption to format your figure captions. They should begin on a
%% new page.

\clearpage

%% No more than seven \figcaption commands are allowed per page,
%% so if you have more than seven captions, insert a \clearpage

%% There must be a \figcaption command for each legend. Key the text of the
%% legend and the optional \label in curly braces. If you wish, you may
%% include the name of the corresponding figure file in square brackets.
%% The label is for identification purposes only. It will not insert the
%% figures themselves into the document.
%% If you want to include your art in the paper, use \plotone.
%% Refer to the on-line documentation for details.

%\figcaption[sgi9259.eps]{This is the first figure and it uses sgi9259.eps as
%its EPS figure file. \label{fig1}}

%% Tables should be submitted one per page, so put a \clearpage before

%% Two options are available to the author for producing tables:  the
%% deluxetable environment provided by the AASTeX package or the LaTeX
%% table environment.  Use of deluxetable is preferred.
%%

%% Three table samples follow, two marked up in the deluxetable environment,
%% one marked up as a LaTeX table.

%% In this first example, note that the \tabletypesize{}
%% command has been used to reduce the font size of the table.
%% Note also that the \label command needs to be placed 

\clearpage
\input{table1.tex}
\clearpage
\input{table2.tex}
\clearpage
\input{table3.tex} 
\clearpage

\clearpage

\clearpage

\clearpage

\begin{figure}[p]
\begin{center}
%\mbox{\includegraphics[width=6in]{Figures/f4.ps}}
\includegraphics[height=8in]{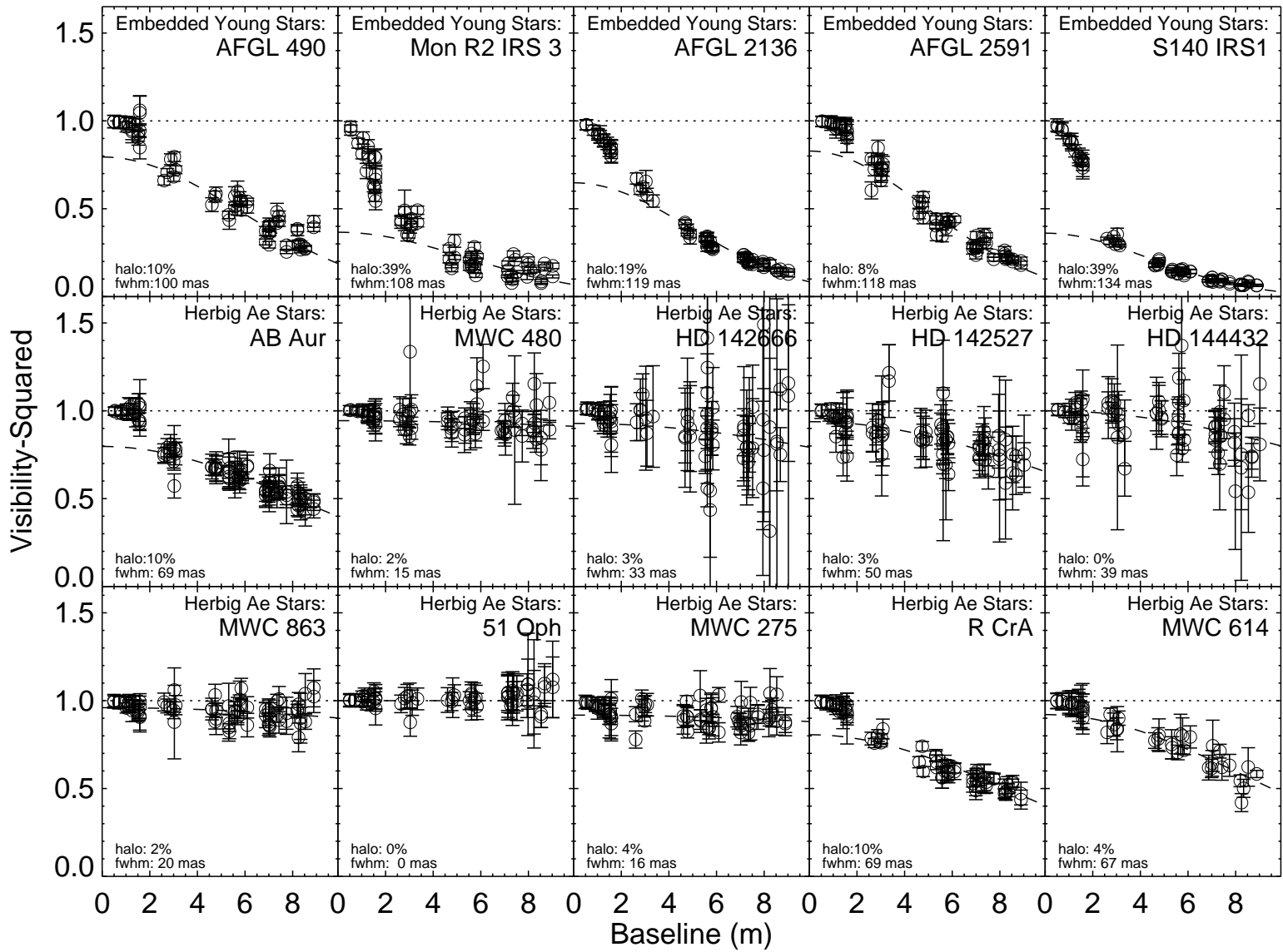}
%\newpage
%\includegraphics[width=6in]{Figures/f4b.ps}
%\clearpage
%\includegraphics[width=6in]{Figures/f4c.ps}
\figcaption{\footnotesize Here we present all the squared-visibility
  data as a function baseline in meters.  
For all stars the observing wavelength was
  10.7$\mu$m.  For this presentation, we present each epoch separately
  and have performed averaging in the (u,v) plane with a smoothing
  length of 0.45~m. Additionally, we have overplotted the best-fit
  circular Gaussian fit and the parameters appear inside each frame
  for reference.  See \S\ref{fitting} for more details on the fitting
  procedure
\label{vis_1d}}
\end{center} 
\end{figure}
\begin{figure}[p]
\begin{center}
\includegraphics[height=8in]{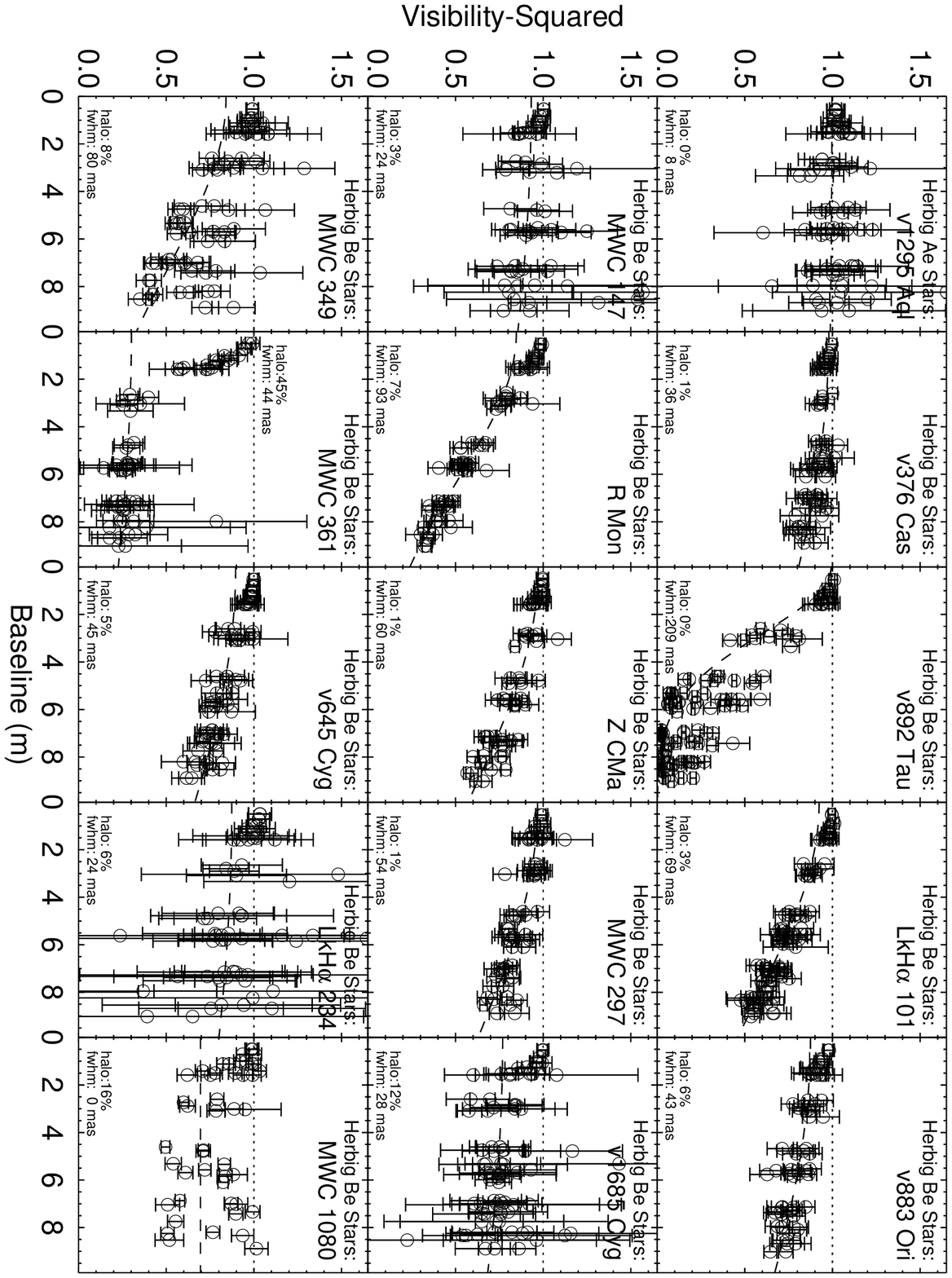}
\\Figure~\ref{vis_1d} (continued)
\end{center}
\end{figure}
\begin{figure}[p]
\begin{center}
\includegraphics[height=8in]{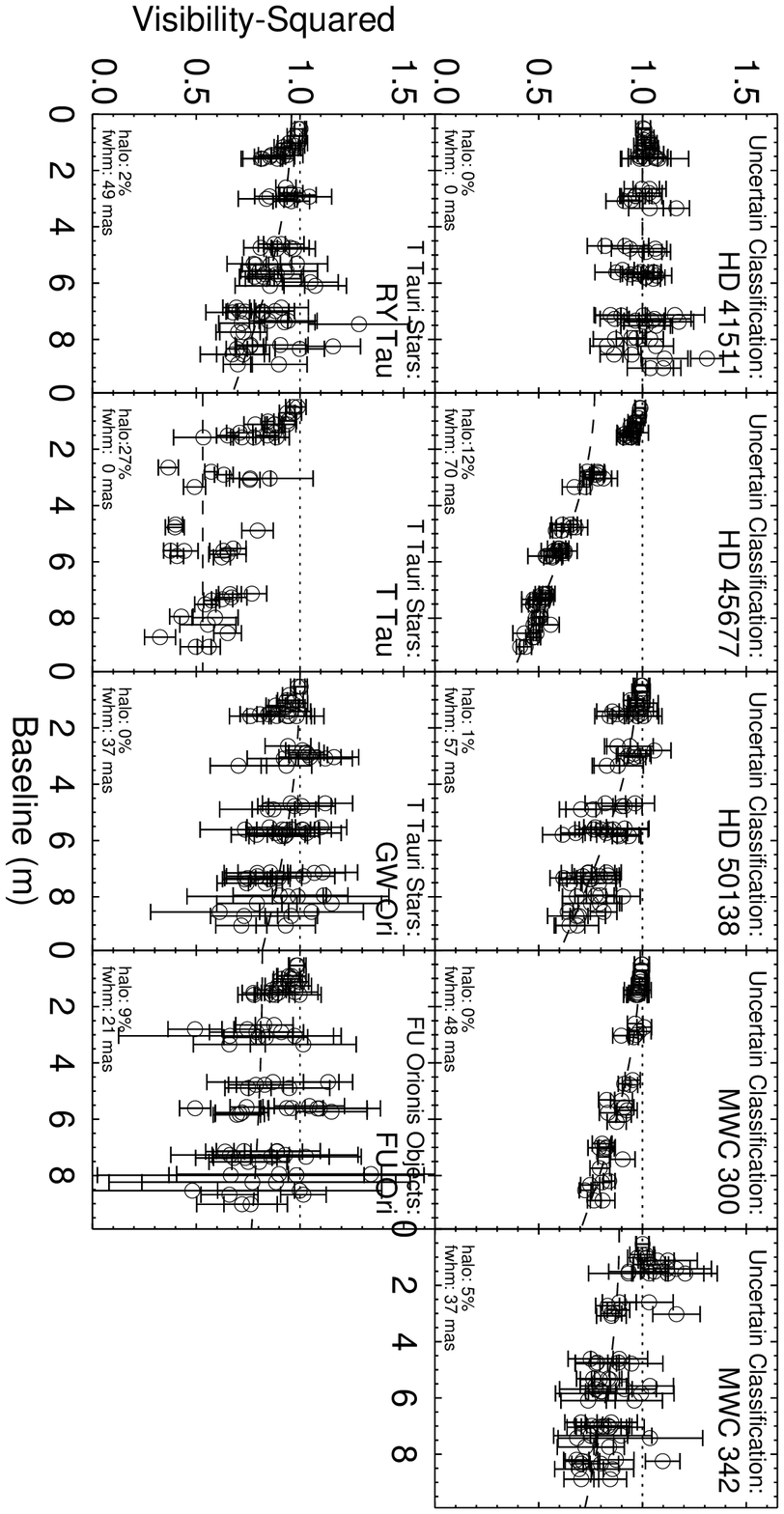}
\\Figure~\ref{vis_1d} (continued)
\end{center}
\end{figure}

\clearpage

\begin{figure}[p]
\begin{center}
%\mbox{\includegraphics[width=6in]{Figures/f4.ps}}
\includegraphics[height=8in]{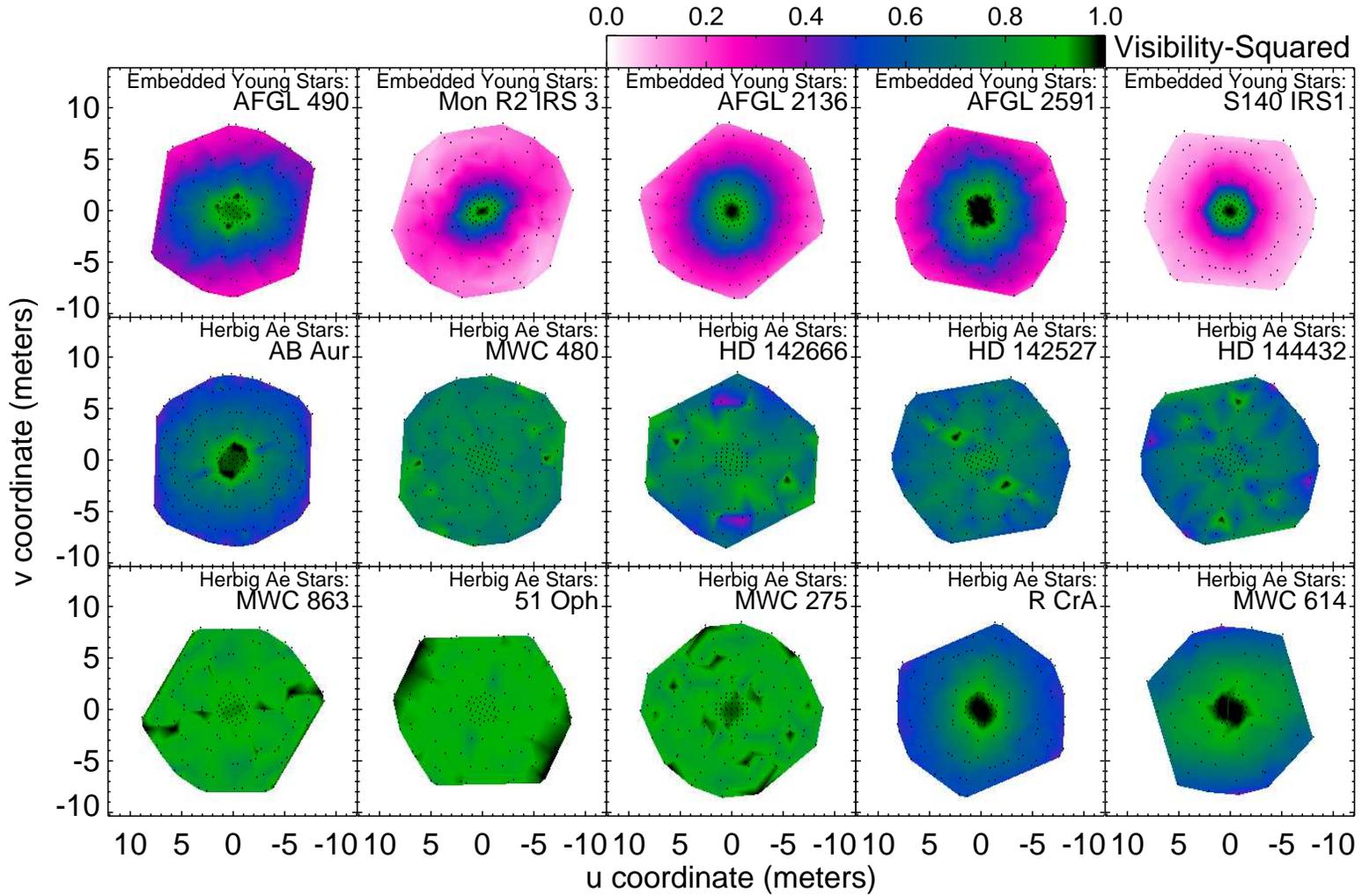}
%\newpage
%\includegraphics[width=6in]{Figures/f4b.ps}
%\clearpage
%\includegraphics[width=6in]{Figures/f4c.ps}
\figcaption{\footnotesize 
Here we present all the squared-visibility data as 2-dimensional function of the (u,v) coordinates.
As for the previous figure, we have averaged on scale of 0.45~meters.  
The visibility data is interpolated onto the grid and shown in color while 
the black dots show the actual locations of the uv datapoints.The intensity scale can found at the top in order to translate the color scale into quantitative values of squared-visibility.
\label{vis_2d}}
\end{center} 
\end{figure}
\begin{figure}[p]
\begin{center}
\includegraphics[height=8in]{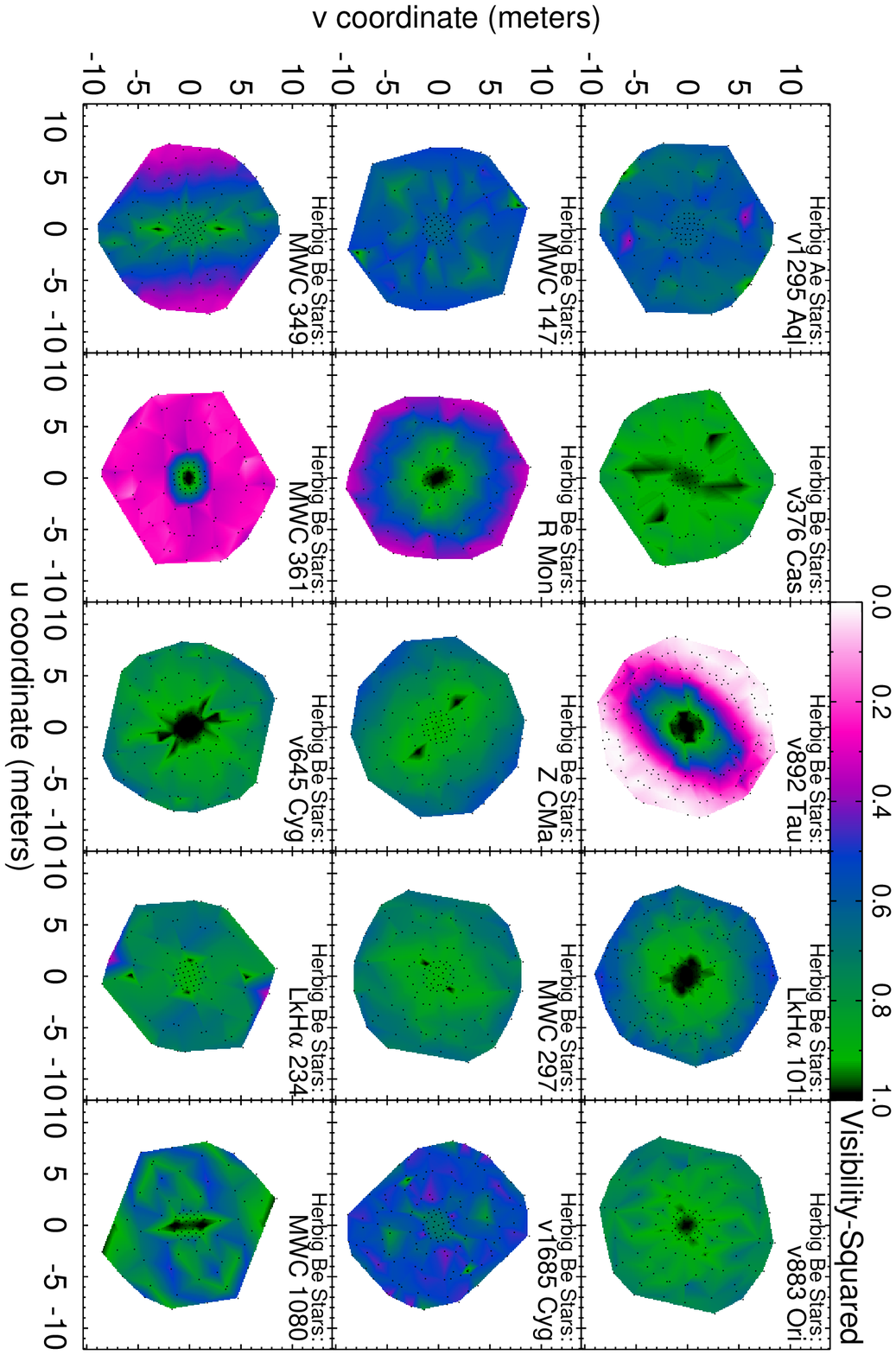}
\\Figure~\ref{vis_2d} (continued)
\end{center}
\end{figure}
\begin{figure}[p]
\begin{center}
\includegraphics[height=8in]{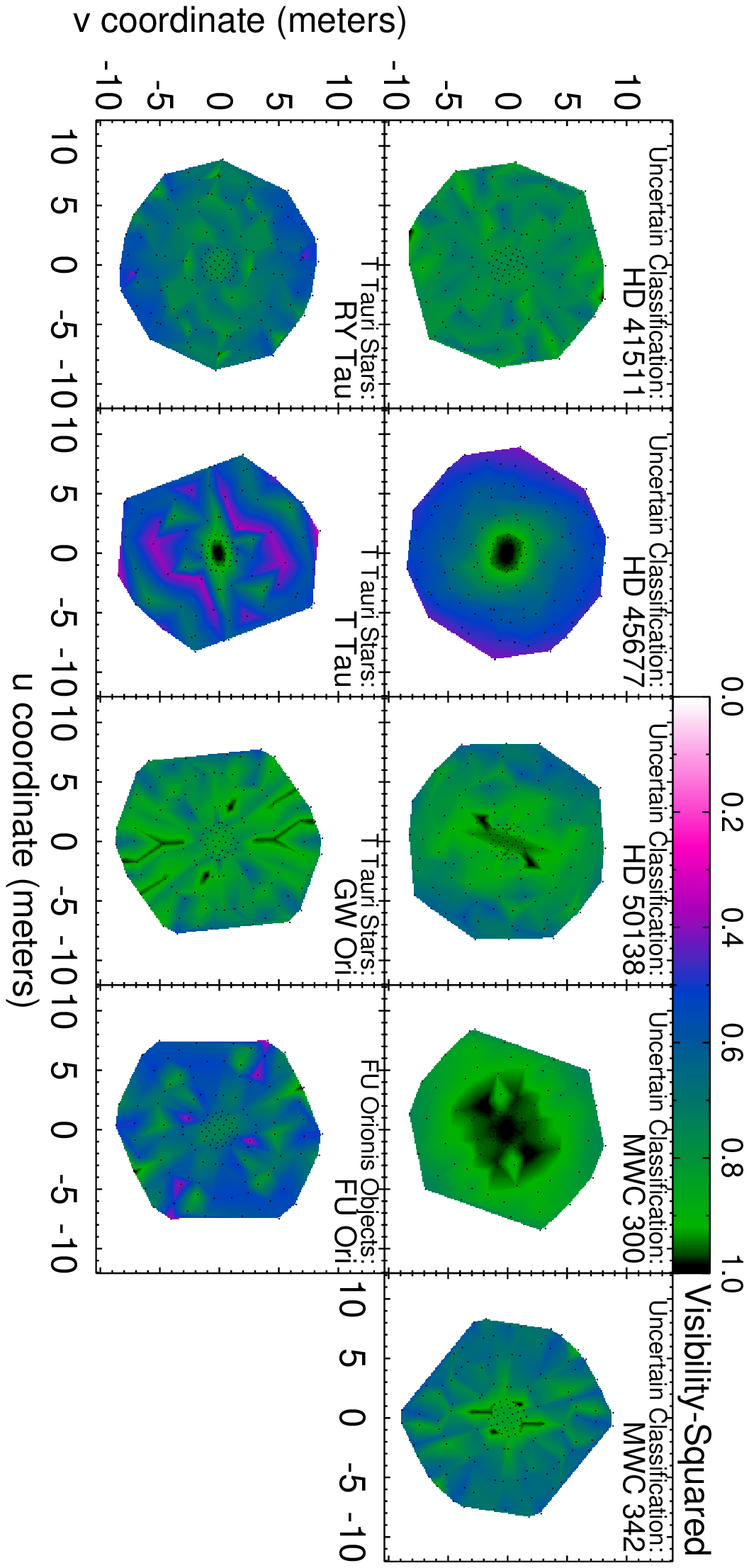}
\\Figure~\ref{vis_2d} (continued)
\end{center}
\end{figure}

\clearpage

\begin{figure}[hbt]
\begin{center}
\includegraphics[angle=90,width=7in]{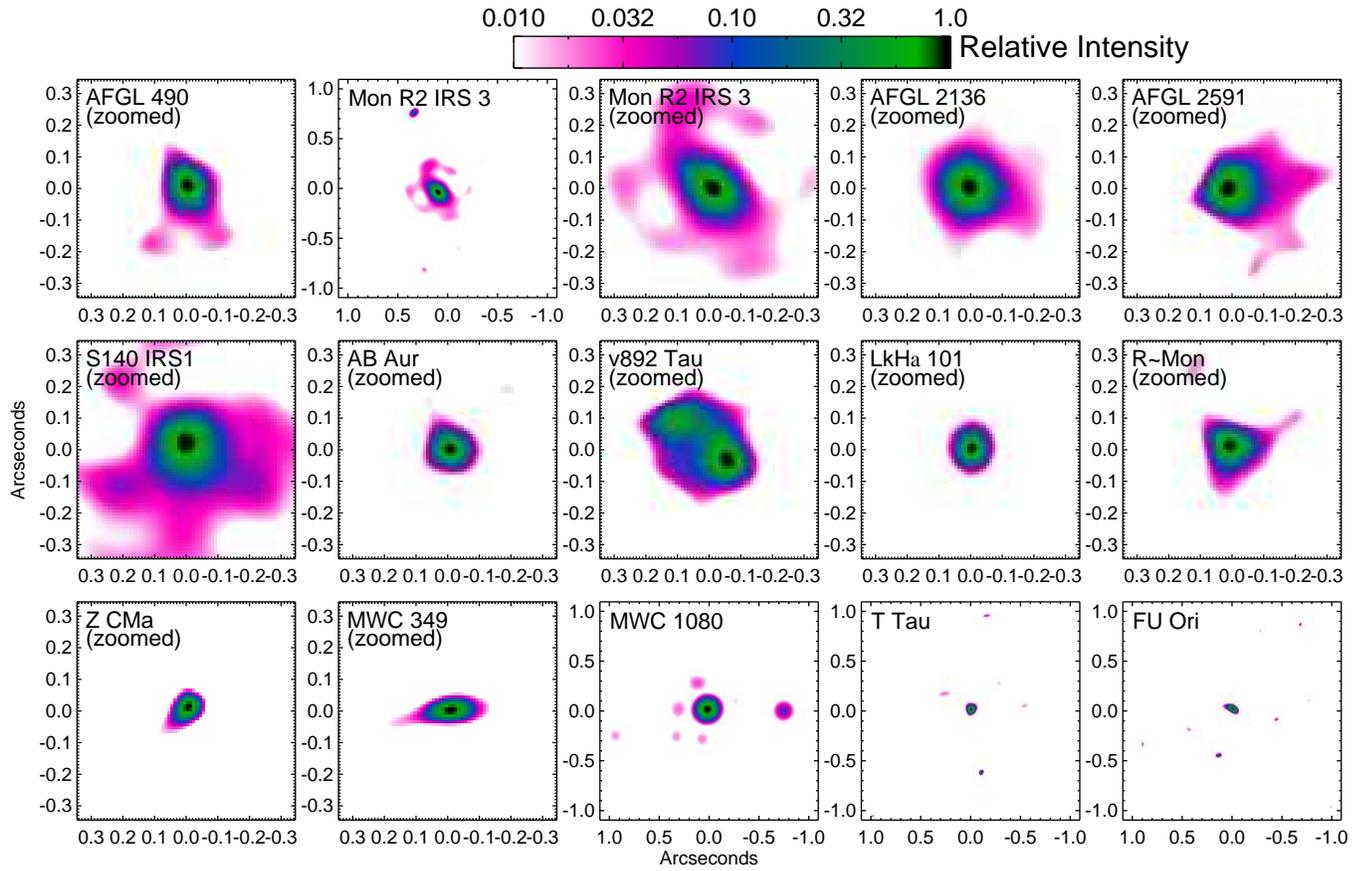}
\figcaption{\footnotesize 
This figure shows the aperture synthesis images of the most resolved targets
in our sample. The intensity scale is logarithmic and is specified in the color bar. 
North is up and East is left in all figures.
\label{images}}
\end{center} 
\end{figure}

\clearpage

\begin{figure}[hbt]
\begin{center}
\includegraphics[angle=90,width=7in]{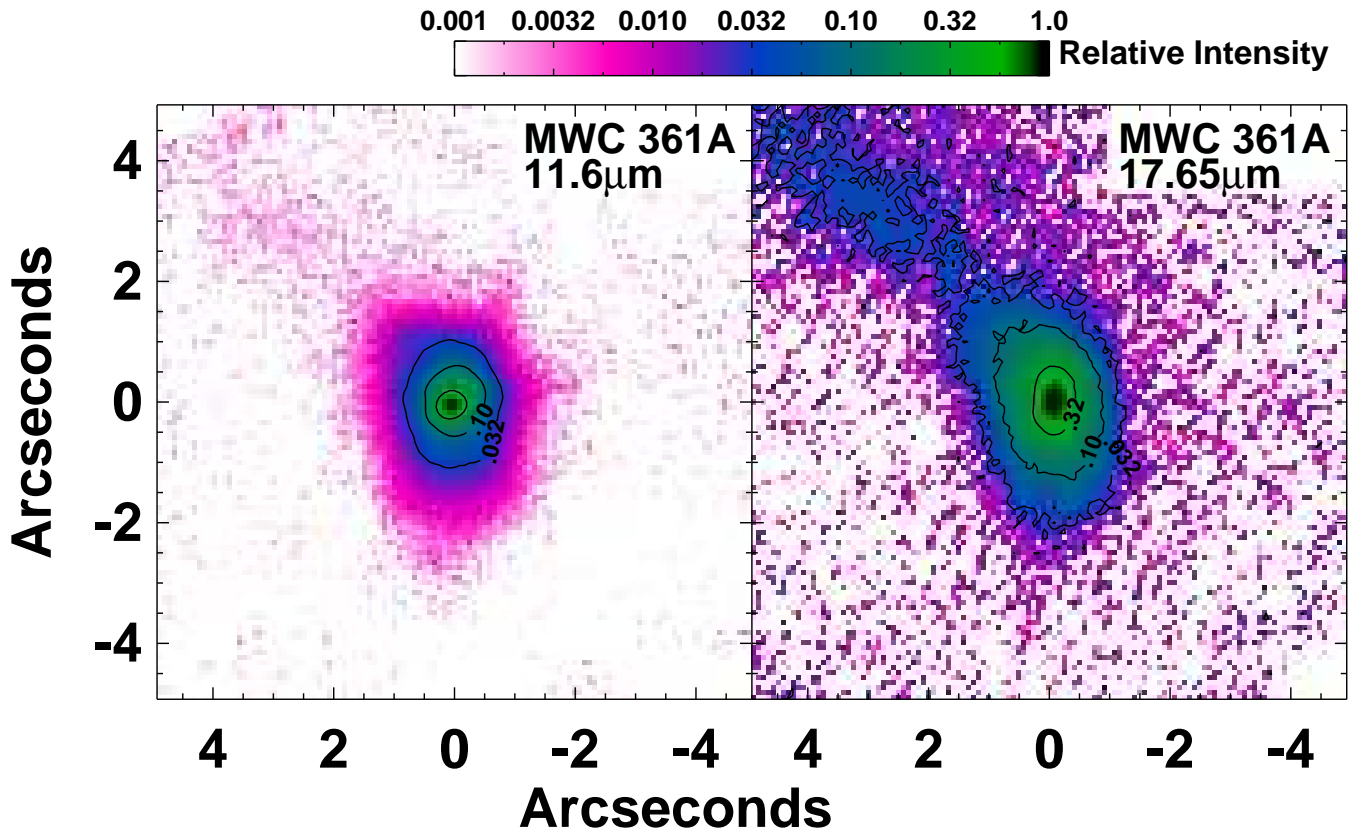}
\figcaption{\footnotesize 
This figure shows imaging of MWC~361A at 11.6$\mu$m and 17.65$\mu$m using the LWS
camera on the Keck telescope without segment-tilting \citep[first presented in][]{perrinthesis}.  
We see an arcsecond-scale
North-South extension which follows the orientation of the underlying
sub-arcsecond binary in this system \citep{monnieriota2006}.  This is a remarkable structure 
suggesting the circumbinary disk is in an advanced stage of  photoevaporation.
The intensity scale is logarithmic and is specified in the color bar. 
North is up and East is left in all figures.
\label{perrin_images}}
\end{center} 
\end{figure}

\clearpage

\begin{figure}[hbt]
\begin{center}
\includegraphics[angle=90,width=7in]{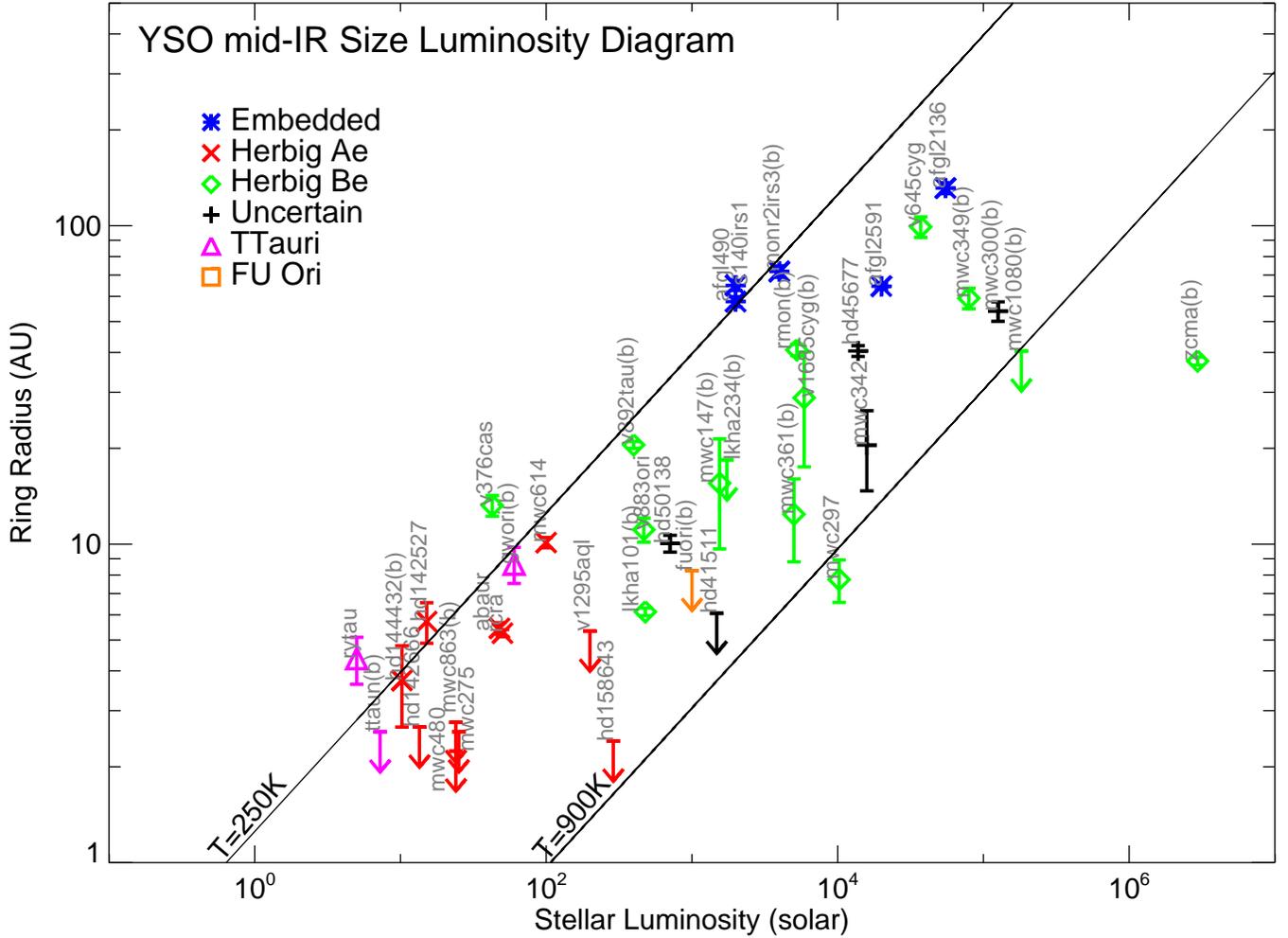}
\figcaption{\footnotesize 
The mid-infrared size -- luminosity diagram.  Different plot symbols correspond to
different YSO types: Embedded YSOs (asterisk), Herbig Ae (X), Herbig Be (diamond), T Tauri (triangle), FU Orionis objects (square), and non-classified emission line objects (plus).   Systems in close binaries are labeled (b).  The lines represent different temperatures for grey dust according to definition in \citet{monnier2002a}.  Objects with FWHM sizes smaller than 35 mas are labeled 
as upper limits.
\label{sizelum}}
\end{center} 
\end{figure}

\clearpage

\begin{figure}[hbt]
\begin{center}
%\mbox{
\includegraphics[angle=90,width=3in]{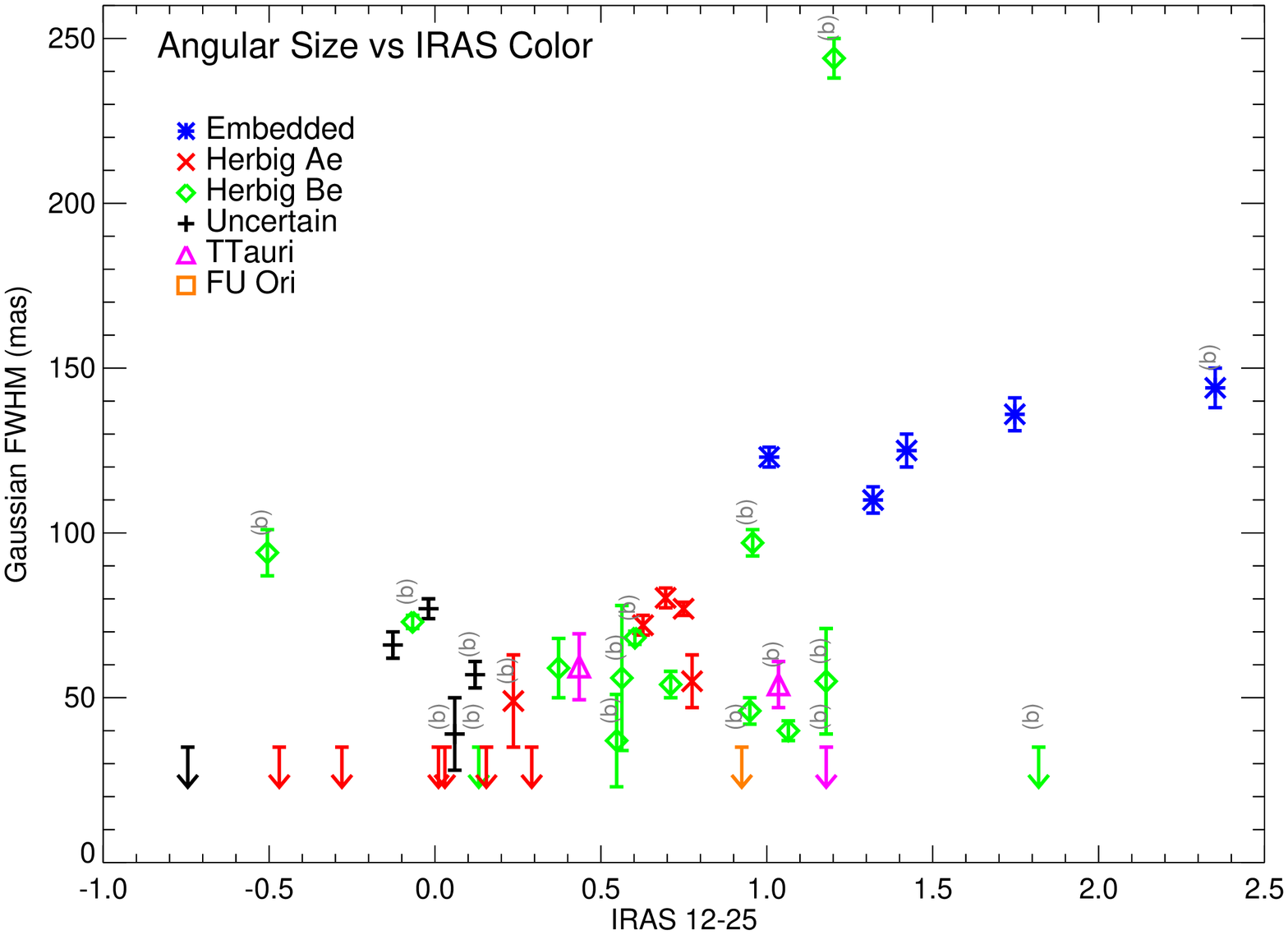}
\hspace{.2in}
\includegraphics[angle=90,width=3in]{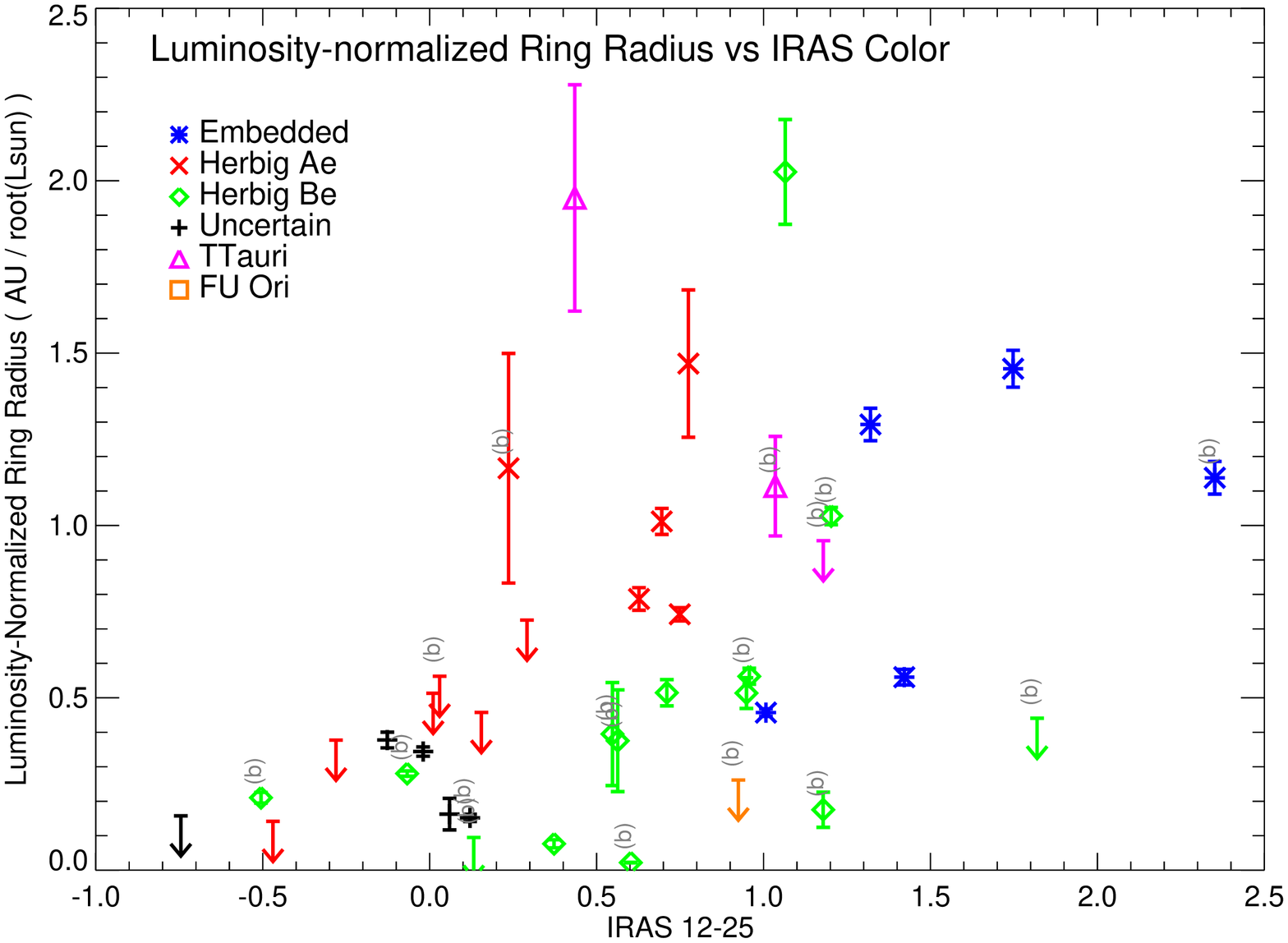}
%}
\figcaption{\footnotesize 
{\em (left panel)} a.
The mid-infrared angular size (Gaussian FWHM) vs. IRAS 12-25 color (-2.5 log F$_\nu$(12$\mu$m) / F$_\nu$(25$\mu$m) ).  Systems in close binaries are labeled (b). 
Objects with FWHM sizes smaller than 35 mas are shown as 35~mas  upper limits.
{\em (right panel)} b. Here we plot the luminosity-normalized ring radius (AU/root luminosity) as a function of
the IRAS 12-25 micron.  The luminosity-normalized ring radius is independent of distance and
is diagnostic of disk structure. Note the significant correlation for the Herbig Ae sample.
\label{sizeiras}}
\end{center} 
\end{figure}

\end{document}

%% file: table1.tex
\clearpage
\begin{deluxetable}{llll}
\tabletypesize{\scriptsize}
\tablecaption{Observing Log
\label{table_obslog}}
\tablehead{
\colhead{Target} & \multicolumn{2}{c}{U.T.} & 
\colhead{Calibrator Names\tablenotemark{a}}\\
                 & \colhead{Date} & \colhead{Time} & }\startdata 
51~Oph               & 2005 May 26 & 10:05 & $\sigma$~Lib                             \\
                     & 2005 May 26 & 10:20 & $\sigma$~Lib, v3879~Sgr                  \\
AB~Aur               & 2004 Aug 30 & 15:12 & $i$~Aur                                  \\
                     & 2004 Aug 30 & 15:34 & $i$~Aur                                  \\
                     & 2004 Aug 31 & 15:29 & $i$~Aur                                  \\
                     & 2004 Sep 01 & 15:37 & $i$~Aur                                  \\
AFGL~2136            & 2005 May 26 & 12:47 & v3879~Sgr                                \\
                     & 2005 May 26 & 13:06 & v3879~Sgr                                \\
AFGL~2591            & 2004 Sep 01 & 09:01 & X~Oph                                    \\
                     & 2004 Sep 01 & 09:28 & X~Oph                                    \\
AFGL~490             & 2004 Sep 01 & 12:23 & $\alpha$~Tau                             \\
                     & 2004 Sep 01 & 12:40 & $\alpha$~Cas                             \\
FU~Ori               & 2005 Feb 20 & 07:28 & $i$~Aur                                  \\
                     & 2005 Feb 20 & 08:33 & $i$~Aur                                  \\
GW~Ori               & 2005 Feb 20 & 07:14 & $i$~Aur                                  \\
                     & 2005 Feb 20 & 08:22 & $\alpha$~Tau, $i$~Aur                    \\
HD~142527            & 2005 May 26 & 08:32 & GM~Lup                                   \\
                     & 2005 May 26 & 09:12 & GM~Lup                                   \\
HD~142666            & 2005 May 26 & 09:49 & $\sigma$~Lib                             \\
                     & 2005 May 26 & 10:11 & $\sigma$~Lib                             \\
HD~144432            & 2005 May 26 & 08:56 & GM~Lup                                   \\
                     & 2005 May 26 & 09:42 & $\sigma$~Lib                             \\
HD~41511             & 2005 Feb 20 & 05:24 & $\alpha$~CMa                             \\
                     & 2005 Feb 20 & 09:21 & $\alpha$~CMa                             \\
HD~45677             & 2005 Feb 19 & 07:32 & $\alpha$~CMa, $\alpha$~Tau               \\
                     & 2005 Feb 19 & 07:32 & $\alpha$~CMa, $\alpha$~Tau               \\
                     & 2005 Feb 19 & 09:08 & $\alpha$~CMa                             \\
HD~50138             & 2005 Feb 20 & 05:50 & $\alpha$~CMa                             \\
                     & 2005 Feb 20 & 09:16 & $\alpha$~CMa                             \\
LkH$\alpha$~101      & 2004 Aug 30 & 14:45 & $\alpha$~Tau, $i$~Aur                    \\
                     & 2004 Aug 31 & 14:56 & $\alpha$~Tau, $i$~Aur                    \\
                     & 2004 Sep 01 & 15:33 & $i$~Aur                                  \\
                     & 2005 Feb 19 & 07:00 & $\alpha$~Tau                             \\
LkH$\alpha$~234      & 2005 May 26 & 14:54 & MO~Cep                                   \\
                     & 2005 May 26 & 15:11 & MO~Cep                                   \\
MWC~1080             & 2004 Sep 01 & 13:25 & $\alpha$~Cas, MO~Cep                     \\
                     & 2004 Sep 01 & 13:28 & $\alpha$~Cas, MO~Cep                     \\
MWC~147              & 2005 Feb 20 & 05:59 & $\alpha$~CMa, $\alpha$~Tau               \\
                     & 2005 Feb 20 & 08:54 & $\alpha$~CMa                             \\
MWC~275              & 2004 Sep 01 & 06:33 & v3879~Sgr                                \\
                     & 2004 Sep 01 & 06:46 & v3879~Sgr                                \\
MWC~297              & 2004 Aug 30 & 09:24 & v3879~Sgr                                \\
                     & 2004 Aug 31 & 07:39 & v3879~Sgr                                \\
MWC~300              & 2004 Aug 31 & 07:44 & v3879~Sgr                                \\
                     & 2004 Aug 31 & 07:55 & v3879~Sgr                                \\
MWC~342              & 2004 Aug 31 & 10:15 & XI~Cyg                                   \\
                     & 2004 Aug 31 & 10:43 & XI~Cyg, v1339~Cyg                        \\
MWC~349              & 2004 Aug 31 & 10:10 & XI~Cyg                                   \\
                     & 2004 Aug 31 & 10:53 & XI~Cyg, v1339~Cyg                        \\
MWC~361              & 2005 May 26 & 14:48 & MO~Cep                                   \\
                     & 2005 May 26 & 15:03 & MO~Cep                                   \\
MWC~480              & 2004 Aug 31 & 15:06 & $i$~Aur                                  \\
                     & 2004 Aug 31 & 15:10 & $i$~Aur                                  \\
                     & 2004 Aug 31 & 15:54 & $\alpha$~Tau, $i$~Aur                    \\
                     & 2004 Aug 31 & 15:58 & $\alpha$~Tau, $i$~Aur                    \\
                     & 2004 Sep 01 & 15:06 & $i$~Aur                                  \\
MWC~614              & 2004 Aug 31 & 09:02 & $\gamma$~Aql                             \\
                     & 2004 Aug 31 & 09:23 & $\gamma$~Aql, R~Lyr                      \\
MWC~863              & 2004 Sep 01 & 05:47 & $\eta$~Sgr, v3879~Sgr                    \\
                     & 2004 Sep 01 & 06:08 & $\eta$~Sgr, v3879~Sgr                    \\
Mon~R2~IRS~3         & 2005 Feb 19 & 07:48 & $\alpha$~CMa                             \\
                     & 2005 Feb 19 & 09:28 & $\alpha$~CMa                             \\
RY~Tau               & 2004 Aug 31 & 15:16 & $\alpha$~Tau                             \\
                     & 2004 Aug 31 & 15:18 & $\alpha$~Tau                             \\
                     & 2004 Aug 31 & 15:49 & $\alpha$~Tau                             \\
                     & 2004 Sep 01 & 15:11 & $i$~Aur                                  \\
R~CrA                & 2004 Aug 31 & 06:16 & $\eta$~Sgr                               \\
                     & 2004 Aug 31 & 06:34 & $\eta$~Sgr                               \\
R~Mon                & 2005 Feb 20 & 05:30 & $\alpha$~CMa                             \\
                     & 2005 Feb 20 & 05:33 & $\alpha$~CMa                             \\
                     & 2005 Feb 20 & 09:03 & $\alpha$~CMa                             \\
S140~IRS1            & 2004 Sep 01 & 11:30 & MO~Cep                                   \\
                     & 2004 Sep 01 & 11:49 & MO~Cep                                   \\
T~Tau                & 2005 Feb 20 & 06:39 & $\alpha$~Tau, $i$~Aur                    \\
                     & 2005 Feb 20 & 07:50 & $\alpha$~Tau, $i$~Aur                    \\
Z~CMa                & 2005 Feb 19 & 07:56 & $\alpha$~CMa                             \\
                     & 2005 Feb 19 & 09:33 & $\alpha$~CMa                             \\
v1295~Aql            & 2005 May 26 & 14:04 & $\gamma$~Aql                             \\
                     & 2005 May 26 & 14:19 & $\gamma$~Aql                             \\
v1685~Cyg            & 2004 Aug 31 & 09:32 & $\gamma$~Aql                             \\
                     & 2004 Aug 31 & 10:58 & XI~Cyg, v1339~Cyg                        \\
                     & 2004 Aug 31 & 11:02 & XI~Cyg, v1339~Cyg                        \\
                     & 2004 Sep 01 & 09:52 & v1339~Cyg                                \\
v376~Cas             & 2004 Sep 01 & 11:25 & $\alpha$~Cas                             \\
                     & 2004 Sep 01 & 11:44 & $\alpha$~Cas                             \\
v645~Cyg             & 2004 Aug 30 & 11:28 & $\beta$~And, $\gamma$~Aql, R~Lyr         \\
                     & 2004 Aug 31 & 10:38 & XI~Cyg, v1339~Cyg                        \\
                     & 2004 Aug 31 & 11:19 & XI~Cyg, v1339~Cyg                        \\
v883~Ori             & 2005 Feb 19 & 07:38 & $\alpha$~Tau                             \\
                     & 2005 Feb 19 & 09:17 & $\alpha$~CMa                             \\
v892~Tau             & 2004 Aug 31 & 14:46 & $\alpha$~Tau, $i$~Aur                    \\
                     & 2004 Aug 31 & 14:51 & $\alpha$~Tau, $i$~Aur                    \\
                     & 2004 Aug 31 & 15:34 & $\alpha$~Tau                             \\
                     & 2004 Aug 31 & 15:39 & $\alpha$~Tau                             \\
                     & 2004 Sep 01 & 14:58 & $\alpha$~Cet                             \\
                     & 2005 Feb 19 & 07:07 & $\alpha$~Tau                             \\
\enddata
\tablenotetext{a}{
All data taken using Pattern~6 (see \S\ref{segmenttilting}). 
All calibrators were assumed to be as unresolved except for
$\alpha$~Tau with UD diameter 19.8~mas \citep{perrin1998}.} 
\end{deluxetable}

%% file: table2.tex
\begin{deluxetable}{llllrrrrrrrl}
\tabletypesize{\scriptsize}
\tablecaption{Source List
\label{table_sources}}
\tablehead{
\colhead{Target} & \colhead{Alternate} & \multicolumn{2}{c}{J2000 Coordinates\tablenotemark{a}}
                 & \colhead{V\tablenotemark{b}} & \colhead{J} & \colhead{H} & \colhead{K} & \colhead{10.7$\mu$m}
                 & \colhead{IRAS12} & \colhead{IRAS25} & \colhead{Binary?}  \\
                 & \colhead{Name}   & \colhead{$\alpha$} & \colhead{$\delta$}
                 & \colhead{} & \colhead{} & \colhead{} & \colhead{} & \colhead{Jy\tablenotemark{c}}
                 & \colhead{Jy\tablenotemark{d}}  & \colhead{Jy} & \colhead{(Ref\tablenotemark{e})} } \startdata
\multicolumn{12}{l}{Embedded Young Stars (Class I Sources)} \\
\hline
            AFGL~490 &                                & 03 27 38.77 & $+$58 47 00.1 &    -- & 10.9 &  8.1 &  5.7 &  58.3$\pm$ 5.8 &  82.4 &  278.0 & N \\
        Mon~R2~IRS~3 &                                & 06 07 47.86 & $-$06 22 56.0 &    -- & 13.2 &  9.8 &  6.6 & 109.9$\pm$11.0 & 470.0 & 4100.0 & Y ( 1) \\
           AFGL~2136 &                                & 18 22 26.38 & $-$13 30 12.0 &    -- & 15.0 & 12.7 &  7.3 &  37.1$\pm$ 3.7 & 155.0 &  574.0 & N \\
           AFGL~2591 &                                & 20 29 24.87 & $+$40 11 19.4 &    -- & 14.3 & 10.8 &  6.6 & 127.7$\pm$12.8 & 439.0 & 1110.0 & N \\
           S140~IRS1 &                                & 22 19 18.28 & $+$63 18 45.8 &    -- & 12.3 &  9.3 &  6.1 & 125.1$\pm$12.5 & 308.0 & 1540.0 & N \\
\hline
\multicolumn{12}{l}{Herbig Ae Stars} \\
\hline
              AB~Aur &                       HD~31293 & 04 55 45.83 & $+$30 33 04.4 &   7.1 &  5.9 &  5.1 &  4.2 &  28.3$\pm$ 2.8 &  27.2 &   48.1 & N \\
             MWC~480 &                       HD~31648 & 04 58 46.26 & $+$29 50 37.1 &   7.7 &  6.9 &  6.3 &  5.5 &  15.4$\pm$ 1.5 &  10.2 &   10.3 & N \\
           HD~142666 &                      v1026~Sco & 15 56 40.02 & $-$22 01 40.0 &   8.8 &  7.4 &  6.7 &  6.1 &   5.9$\pm$ 0.6 &   8.6 &   11.2 & N \\
           HD~142527 &                                & 15 56 41.89 & $-$42 19 23.3 &   8.3 &  6.5 &  5.7 &  5.0 &  11.1$\pm$ 1.1 &  10.4 &   21.2 & N \\
           HD~144432 &                                & 16 06 57.95 & $-$27 43 09.4 &   8.2 &  7.1 &  6.5 &  5.9 &   9.8$\pm$ 1.0 &   7.5 &    9.4 & Y ( 2) \\
             MWC~863 &                      HD~150193 & 16 40 17.92 & $-$23 53 45.2 &   8.9 &  6.9 &  6.2 &  5.5 &  18.9$\pm$ 3.8 &  17.6 &   18.1 & Y ( 3) \\
              51~Oph &                      HD~158632 & 17 31 24.97 & $-$23 57 45.3 &   4.8 &  4.9 &  4.7 &  4.3 &  16.9$\pm$ 1.7 &  15.7 &   10.2 & N \\
             MWC~275 &                      HD~163296 & 17 56 21.29 & $-$21 57 21.8 &   6.9 &  6.2 &  5.5 &  4.8 &  15.3$\pm$ 1.5 &  18.2 &   21.0 & N \\
               R~CrA &                                & 19 01 53.68 & $-$36 57 08.2 &  11.5 &  6.9 &  5.0 &  2.9 & 110.8$\pm$11.1 & 111.0 &  222.0 & N \\
             MWC~614 &                      HD~179218 & 19 11 11.24 & $+$15 47 15.6 &   7.2 &  7.0 &  6.6 &  6.0 &  20.6$\pm$ 4.0 &  23.4 &   43.6 & N \\
           v1295~Aql &                      HD~190973 & 20 03 02.51 & $+$05 44 16.7 &   7.8 &  7.2 &  6.6 &  5.9 &   6.8$\pm$ 0.7 &   7.2 &    5.5 & N \\
\hline
\multicolumn{12}{l}{Herbig Be Stars} \\
\hline
            v376~Cas &                                & 00 11 26.52 & $+$58 50 03.7 &  15.6 & 11.0 &  8.4 &  6.3 &  21.1$\pm$ 2.1 &  33.0 &   88.1 & N \\
            v892~Tau &                        Elias~1 & 04 18 40.62 & $+$28 19 15.5 &  15.3 &  8.7 &  7.0 &  5.8 &  34.1$\pm$ 3.9 &  33.2 &   99.9 & Y ( 4) \\
     LkH$\alpha$~101 &                                & 04 30 14.44 & $+$35 16 24.0 &  15.7 &  8.3 &  5.6 &  3.0 & 289.7$\pm$29.0 & 362.0 &  340.0 & Y ( 5) \\
            v883~Ori &                IRAS~05358-0704 & 05 38 18.10 & $-$07 02 25.9 &  14.4 &  9.3 &  6.8 &  5.2 &  39.8$\pm$ 6.0 &  52.5 &  127.0 & N \\
             MWC~147 &                      HD~259431 & 06 33 05.20 & $+$10 19 19.9 &   8.8 &  7.5 &  6.7 &  5.7 &  10.0$\pm$ 1.0 &  12.5 &   20.2 & Y ( 6) \\
               R~Mon &                                & 06 39 09.95 & $+$08 44 09.7 &  10.4 &  9.7 &  8.0 &  6.4 &  42.8$\pm$ 4.3 &  54.7 &  132.0 & Y ( 7) \\
               Z~CMa &                                & 07 03 43.17 & $-$11 33 06.3 &   9.9 &  6.5 &  5.2 &  3.8 & 144.7$\pm$14.5 & 127.0 &  221.0 & Y ( 8) \\
             MWC~297 &                         NZ~Ser & 18 27 39.53 & $-$03 49 52.1 &  12.3 &  6.1 &  4.4 &  3.0 &  84.7$\pm$ 8.5 & 159.0 &  224.0 & N \\
           v1685~Cyg &                        MWC~340 & 20 20 28.25 & $+$41 21 51.5 &  10.7 &  7.9 &  6.8 &  5.8 &   4.6$\pm$ 0.8 & -- & -- & Y ( 9) \\
             MWC~349 &                      v1478~Cyg & 20 32 45.53 & $+$40 39 36.6 &  13.2 &  6.2 &  4.8 &  3.2 & 153.9$\pm$37.9 & 179.0 &  112.0 & Y (10) \\
             MWC~361 &                      HD~200775 & 21 01 36.91 & $+$68 09 47.7 &   7.4 &  6.1 &  5.5 &  4.7 &   8.2$\pm$ 0.8 &  26.7 &   76.8 & Y (11) \\
            v645~Cyg &                      AFGL~2789 & 21 39 58.25 & $+$50 14 20.9 &  13.6 & 10.9 &  9.2 &  6.8 &  57.4$\pm$12.0 & 114.0 &  219.0 & N \\
     LkH$\alpha$~234 &                                & 21 43 06.82 & $+$66 06 54.2 &  11.9 &  9.5 &  8.2 &  7.1 &   3.9$\pm$ 0.4 &  14.8 &   79.0 & Y (12) \\
            MWC~1080 &                       v628~Cas & 23 17 25.59 & $+$60 50 43.6 &  11.6 &  7.5 &  6.0 &  4.8 &  15.5$\pm$ 1.8 &  22.2 &   25.1 & Y ( 3) \\
\hline
\multicolumn{12}{l}{Emission-line Stars of Uncertain Classification (possible Herbigs, B[e] stars, or other)} \\
\hline
            HD~41511 &                                & 06 04 59.13 & $-$16 29 03.9 &   5.0 &  2.9 &  2.1 &  1.7 & 208.7$\pm$20.9 & 144.0 &   72.2 & N \\
            HD~45677 &                                & 06 28 17.42 & $-$13 03 11.0 &   8.1 &  7.2 &  6.3 &  4.8 & 181.9$\pm$28.1 & 146.0 &  143.0 & N \\
            HD~50138 &                                & 06 51 33.41 & $-$06 57 59.2 &   6.6 &  5.9 &  5.1 &  4.1 &  94.6$\pm$ 9.5 &  70.3 &   62.5 & N \\
             MWC~300 &                       v431~Sct & 18 29 25.70 & $-$06 04 37.2 &  10.5 &  9.3 &  8.2 &  6.2 &  64.6$\pm$ 6.5 & 101.0 &  113.0 & Y ( 3) \\
             MWC~342 &                      v1972~Cyg & 20 23 03.61 & $+$39 29 49.9 &  10.6 &  7.0 &  5.8 &  4.7 &  51.5$\pm$13.5 &  46.2 &   48.6 & N \\
\hline
\multicolumn{12}{l}{T Tauri Stars} \\
\hline
              RY~Tau &                                & 04 21 57.40 & $+$28 26 35.5 &  10.2 &  7.2 &  6.1 &  5.4 &  14.4$\pm$ 1.4 &  17.5 &   26.1 & N \\
               T~Tau &                                & 04 21 59.43 & $+$19 32 06.4 &   9.6 &  7.2 &  6.2 &  5.3 &  10.7$\pm$ 1.1 &  14.9 &   44.2 & Y (13) \\
              GW~Ori &                                & 05 29 08.39 & $+$11 52 12.6 &   9.9 &  7.7 &  7.1 &  6.6 &  10.0$\pm$ 1.0 &   7.9 &   20.5 & Y (14) \\
\hline
\multicolumn{12}{l}{FU Orionis Objects} \\
\hline
              FU~Ori &                                & 05 45 22.36 & $+$09 04 12.4 &   8.9 &  6.5 &  5.7 &  5.2 &   5.9$\pm$ 0.6 &   5.9 &   14.1 & Y (15)
\enddata
\tablenotetext{a}{Coordinates and JHK magnitudes are from 2MASS \citep{2mass}.} 
\tablenotetext{b}{V band magnitudes from SIMBAD.} 
\tablenotetext{c}{Flux included within aperture of diameter of 3\asec and was derived from this work.} 
\tablenotetext{d}{IRAS12 and IRAS25 fluxes are  from IRAS point source catalog \citep{iras} using broadband filters from 8-15$\mu$m and 15-30$\mu$m.} 
\tablenotetext{e}{Binary references:       1. \citet{koresko1993}.        2. \citet{carmona2007}.        3. \citet{corporon1999}.        4. \citet{smith2005}.        5. \citet{tuthill2001}.        6. \citet{brandvigAAS_2007}.        7. \citet{thomasIAU_2006}.        8. \citet{rmgzcma2002}.        9. \citet{corderoAAS_2007}.       10. \citet{cohen1985}.       11. \citet{monnieriota2006}.       12. \citet{leinert1997}.       13. \citet{dyck1982}.       14. \citet{mathieu1991}.       15. \citet{wang2004}.  }
\end{deluxetable}

%% file: table3.tex
\begin{deluxetable}{l|rrr|rrrrr|l}
\tabletypesize{\scriptsize}
\tablecaption{1-D and 2-D Characteristic Sizes at 10.7$\mu$m
\label{table_sizes}}
\tablehead{
\colhead{Target} & \multicolumn{3}{c}{1-D Gaussian\tablenotemark{a}} & \multicolumn{5}{c}{2-D Gaussian\tablenotemark{b}} & Comments \\
                 & \colhead{FWHM} & \colhead{Halo} & \colhead{$\chi^2_\nu$} 
&  \multicolumn{2}{c}{FWHM (mas)} & \colhead{PA} & \colhead{Halo} & \colhead{$\chi^2_\nu$} & \\
                 & \colhead{mas}  & \colhead{(\%)}  & 
& \colhead{Major} & \colhead{Minor} & \colhead{deg} & \colhead{(\%)} &  & }
\startdata
\multicolumn{10}{l}{Embedded Young Stars (Class I Sources)} \\
\hline
            AFGL~490 & 100$\pm$ 5 & 11$\pm$3 &  7.0 & 110$\pm$ 4 &  81$\pm$ 3 &  20$\pm$  2 & 12$\pm$2 &  2.4 & Faint extension to South
                  \\
        Mon~R2~IRS~3 & 108$\pm$16 & 39$\pm$6 & 20.0 & 144$\pm$ 6 &  88$\pm$ 5 &  41$\pm$  2 & 35$\pm$2 &  6.1 & Nebulosity and binary at $\rho=836$~mas, $\theta=16.7\arcdegg$
                  \\
           AFGL~2136 & 120$\pm$ 3 & 20$\pm$2 &  2.1 & 125$\pm$ 5 & 115$\pm$ 3 &  44$\pm$ 14 & 19$\pm$2 &  1.7 & Faint extension to West-Southwest
                  \\
           AFGL~2591 & 118$\pm$ 3 &  9$\pm$4 &  4.2 & 123$\pm$ 3 & 111$\pm$ 3 & 114$\pm$  7 &  9$\pm$4 &  3.4 & Nebulosity to West
                  \\
           S140~IRS1 & 134$\pm$ 3 & 40$\pm$1 &  5.5 & 136$\pm$ 5 & 133$\pm$ 4 & 117$\pm$ 25 & 40$\pm$2 &  5.6 & Extensive nebulosity to South
                  \\
\hline
\multicolumn{10}{l}{Herbig Ae Stars} \\
\hline
              AB~Aur &  70$\pm$ 3 & 11$\pm$1 &  0.4 &  72$\pm$ 3 &  68$\pm$ 6 &  27$\pm$ 25 & 11$\pm$1 &  0.4 & Slight extension PA$\sim$30$\arcdegg$ but mostly symmetrical
                  \\
             MWC~480 &  16$\pm$20 &  3$\pm$2 &  0.8 &  29$\pm$24 &   0$\pm$17 &  29$\pm$ 32 &  2$\pm$2 &  0.8 &  
                  \\
           HD~142666 &  33$\pm$24 &  4$\pm$2 &  0.6 &  65$\pm$65 &   0$\pm$40 & 173$\pm$ 56 &  2$\pm$4 &  0.5 &  
                  \\
           HD~142527 &  50$\pm$ 6 &  3$\pm$1 &  0.4 &  55$\pm$ 8 &  45$\pm$ 4 &  60$\pm$ 22 &  3$\pm$1 &  0.4 &  
                  \\
           HD~144432 &  39$\pm$ 5 &  0$\pm$1 &  1.0 &  49$\pm$14 &  25$\pm$25 & 103$\pm$ 31 &  0$\pm$1 &  0.9 &  
                  \\
             MWC~863 &  21$\pm$21 &  2$\pm$2 &  1.0 &  33$\pm$ 6 &   8$\pm$14 &  12$\pm$ 21 &  2$\pm$1 &  0.9 &  
                  \\
              51~Oph &   0$\pm$ 4 &  0$\pm$1 &  0.3 &   0$\pm$27 &   0$\pm$ 1 &  93$\pm$ 53 &  0$\pm$1 &  0.3 &  
                  \\
             MWC~275 &  17$\pm$16 &  4$\pm$3 &  1.2 &  24$\pm$24 &   8$\pm$ 9 &  60$\pm$ 51 &  4$\pm$3 &  1.2 &  
                  \\
               R~CrA &  69$\pm$ 2 & 10$\pm$1 &  0.9 &  77$\pm$ 2 &  64$\pm$ 2 &  78$\pm$  4 & 10$\pm$1 &  0.6 &  
                  \\
             MWC~614 &  68$\pm$ 4 &  4$\pm$1 &  0.7 &  80$\pm$ 3 &  52$\pm$ 8 &  17$\pm$  9 &  4$\pm$1 &  0.3 &  
                  \\
           v1295~Aql &   8$\pm$15 &  0$\pm$1 &  0.5 &  34$\pm$34 &   0$\pm$ 1 &  22$\pm$ 49 &  0$\pm$1 &  0.5 &  
                  \\
\hline
\multicolumn{10}{l}{Herbig Be Stars} \\
\hline
            v376~Cas &  37$\pm$ 3 &  1$\pm$2 &  0.5 &  40$\pm$ 3 &  33$\pm$ 5 & 134$\pm$ 25 &  1$\pm$2 &  0.5 &  
                  \\
            v892~Tau & 210$\pm$ 8 &  0$\pm$3 & 21.6 & 244$\pm$ 6 & 123$\pm$ 9 &  49$\pm$  1 &  0$\pm$2 &  2.1 & Resolved circumbinary disk
                  \\
     LkH$\alpha$~101 &  70$\pm$ 3 &  4$\pm$2 &  0.8 &  73$\pm$ 2 &  62$\pm$ 9 & 180$\pm$ 14 &  4$\pm$2 &  0.6 & Slight extension North-South
                  \\
            v883~Ori &  43$\pm$ 4 &  6$\pm$2 &  0.9 &  46$\pm$ 4 &  37$\pm$10 & 110$\pm$ 31 &  7$\pm$3 &  0.9 &  
                  \\
             MWC~147 &  24$\pm$14 &  3$\pm$5 &  0.6 &  37$\pm$14 &   0$\pm$11 &  52$\pm$ 31 &  3$\pm$7 &  0.6 &  
                  \\
               R~Mon &  94$\pm$ 2 &  8$\pm$1 &  0.7 &  97$\pm$ 4 &  91$\pm$ 2 &  66$\pm$ 23 &  8$\pm$1 &  0.7 & Mostly circularly symmetric
                  \\
               Z~CMa &  61$\pm$ 4 &  1$\pm$1 &  2.4 &  68$\pm$ 2 &  41$\pm$ 6 & 137$\pm$  5 &  2$\pm$1 &  0.6 & Faint extension toward companion along PA$\sim$128$\arcdegg$
                  \\
             MWC~297 &  55$\pm$ 6 &  2$\pm$1 &  1.2 &  59$\pm$ 9 &  50$\pm$ 6 &  40$\pm$ 21 &  2$\pm$1 &  1.1 &  
                  \\
           v1685~Cyg &  29$\pm$17 & 12$\pm$6 &  0.6 &  56$\pm$23 &  19$\pm$19 & 112$\pm$ 26 & 11$\pm$6 &  0.5 &  
                  \\
             MWC~349 &  81$\pm$ 6 &  8$\pm$3 &  4.0 &  94$\pm$ 7 &  26$\pm$15 &  95$\pm$  6 &  8$\pm$3 &  1.1 & Symmetric elongation
                  \\
             MWC~361 &  45$\pm$17 & 45$\pm$2 &  0.3 &  55$\pm$14 &  39$\pm$30 &  17$\pm$ 26 & 45$\pm$2 &  0.3 & Arcsec-scale halo elongated North-South. See Figure~7.
                  \\
            v645~Cyg &  46$\pm$11 &  5$\pm$4 &  1.0 &  55$\pm$ 4 &  38$\pm$19 &  34$\pm$ 16 &  5$\pm$4 &  0.9 &  
                  \\
     LkH$\alpha$~234 &  27$\pm$19 &  6$\pm$1 &  0.2 &  42$\pm$42 &   0$\pm$44 &  91$\pm$ 34 &  6$\pm$2 &  0.2 &  
                  \\
            MWC~1080 &   0$\pm$10 & 17$\pm$3 &  8.1 &   0$\pm$44 &   0$\pm$ 1 & 115$\pm$ 49 & 17$\pm$2 &  8.4 & Binary: $\rho=764$~mas, $\theta=-91.1\arcdegg$
                  \\
\hline
\multicolumn{10}{l}{Emission-line Stars of Uncertain Classification (possible Herbigs, B[e] stars, or other)} \\
\hline
            HD~41511 &   0$\pm$ 6 &  0$\pm$2 &  1.6 &  19$\pm$26 &   9$\pm$ 9 &  36$\pm$ 90 &  0$\pm$2 &  1.7 &  
                  \\
            HD~45677 &  70$\pm$ 3 & 12$\pm$1 &  1.0 &  77$\pm$ 3 &  66$\pm$ 3 &  70$\pm$  5 & 12$\pm$1 &  0.7 &  
                  \\
            HD~50138 &  58$\pm$ 6 &  1$\pm$1 &  1.4 &  66$\pm$ 4 &  46$\pm$ 9 &  63$\pm$  6 &  1$\pm$1 &  1.0 &  
                  \\
             MWC~300 &  49$\pm$ 3 &  1$\pm$1 &  0.6 &  57$\pm$ 4 &  44$\pm$ 2 &  35$\pm$ 14 &  0$\pm$1 &  0.4 &  
                  \\
             MWC~342 &  38$\pm$22 &  6$\pm$1 &  1.0 &  39$\pm$11 &  37$\pm$37 & 107$\pm$ 90 &  6$\pm$2 &  1.0 &  
                  \\
\hline
\multicolumn{10}{l}{T Tauri Stars} \\
\hline
              RY~Tau &  49$\pm$ 4 &  2$\pm$1 &  0.9 &  59$\pm$10 &  37$\pm$25 &  12$\pm$ 45 &  2$\pm$3 &  0.8 &  
                  \\
               T~Tau &   0$\pm$13 & 27$\pm$2 &  3.5 &  29$\pm$21 &   0$\pm$ 7 &   1$\pm$ 31 & 26$\pm$2 &  3.6 & Binary: $\rho=639$~mas, $\theta=-170.9\arcdegg$
                  \\
              GW~Ori &  37$\pm$20 &  0$\pm$2 &  0.7 &  54$\pm$ 7 &  25$\pm$29 & 107$\pm$ 10 &  0$\pm$5 &  0.6 &  
                  \\
\hline
\multicolumn{10}{l}{FU Orionis Objects} \\
\hline
              FU~Ori &  21$\pm$22 & 10$\pm$5 &  1.2 &  55$\pm$27 &   0$\pm$13 &  65$\pm$ 35 &  8$\pm$4 &  1.2 & Binary: $\rho=488$~mas, $\theta=163.3\arcdegg$
                    
\enddata
\tablenotetext{a}{Parameters of 1-dimensional Gaussian fits: Full-width at half-maximum (FWHM) in milliarcseconds (mas), percentage of light coming from ``halo'' on scales larger than $\sim$0.5\arcsecc. } 
\tablenotetext{b}{Parameters of 2-dimensional Gaussian fits: Full-width at half-maximum (FWHM) along major and minor axes respectively, Position angle (PA) of major axis in degrees East of North, percentage of light coming from extended ``halo.''}
{}
\end{deluxetable}